\documentclass[11pt]{article}

\usepackage{amssymb}
\usepackage{amsmath}
\usepackage{MnSymbol}
\usepackage{hyperref}
\numberwithin{equation}{section}

\usepackage{graphicx}
	\usepackage{epstopdf}
\makeatletter
\long\def\UR#1{\leavevmode\setbox\@tempboxa\hbox{#1}\@tempdima\fboxrule
    \advance\@tempdima \fboxsep \advance\@tempdima \dp\@tempboxa
   \hbox{\lower \@tempdima\hbox
  {\vbox{\hrule \@height \fboxrule
          \hbox{  \hskip\fboxsep
          \vbox{\vskip\fboxsep \box\@tempboxa\vskip\fboxsep}\hskip
                 \fboxsep\vrule \@width \fboxrule}%
                  }}}}
\makeatother

\makeatletter
\long\def\LR#1{\leavevmode\setbox\@tempboxa\hbox{#1}\@tempdima\fboxrule
    \advance\@tempdima \fboxsep \advance\@tempdima \dp\@tempboxa
   \hbox{\lower \@tempdima\hbox
  {\vbox{
          \hbox{  \hskip\fboxsep
          \vbox{\vskip\fboxsep \box\@tempboxa\vskip\fboxsep}\hskip
                 \fboxsep\vrule \@width \fboxrule}%
                 \hrule \@height \fboxrule}}}}
\makeatother

\makeatletter
\long\def\UL#1{\leavevmode\setbox\@tempboxa\hbox{#1}\@tempdima\fboxrule
    \advance\@tempdima \fboxsep \advance\@tempdima \dp\@tempboxa
   \hbox{\lower \@tempdima\hbox
  {\vbox{\hrule \@height \fboxrule
          \hbox{\vrule \@width \fboxrule \hskip\fboxsep
          \vbox{\vskip\fboxsep \box\@tempboxa\vskip\fboxsep}\hskip
                 \fboxsep }%
                  }}}}
\makeatother

\makeatletter
\long\def\LL#1{\leavevmode\setbox\@tempboxa\hbox{#1}\@tempdima\fboxrule
    \advance\@tempdima \fboxsep \advance\@tempdima \dp\@tempboxa
   \hbox{\lower \@tempdima\hbox
  {\vbox{
          \hbox{\vrule \@width \fboxrule \hskip\fboxsep
          \vbox{\vskip\fboxsep \box\@tempboxa\vskip\fboxsep}\hskip
                 \fboxsep }%
                 \hrule \@height \fboxrule}}}}
\makeatother

\title{Process, Distinction, Groupoids and Clifford Algebras: an Alternative View of the Quantum Formalism. }
\author{B. J. Hiley.}
\date{TPRU, Birkbeck, University of London, Malet Street, \\London WC1E 7HX.}
\begin{document}
\maketitle

\begin{abstract}
In this paper we start from a basic notion of process,  which we structure into two groupoids, one orthogonal and one symplectic.  By introducing additional structure, we convert these groupoids into orthogonal and symplectic Clifford algebras respectively.  We show how the orthogonal Clifford algebra, which include the Schr\"{o}dinger, Pauli and Dirac formalisms, describe the {\em classical} light-cone structure of space-time, as well as providing a basis for the description of quantum phenomena.  By constructing an orthogonal Clifford bundle with a Dirac connection, we make contact with quantum mechanics through the Bohm formalism which emerges quite naturally from the connection, showing that it is a structural feature of the mathematics.  We then generalise the approach to include the symplectic Clifford algebra, which leads us to a non-commutative geometry with projections onto shadow manifolds.  These shadow manifolds are none other than examples of the phase space constructed by Bohm.  We also argue that this provides us with a mathematical structure that fits the implicate-explicate order proposed by Bohm.
\end{abstract}
 

\section{The Algebra of Process.}

	Traditionally basic theories of quantum phenomena are described in terms of the dynamical properties of particles-in-interaction, or more basically, fields-in-interaction built on an  {\em a priori} given manifold.  Special relativity demands this manifold is a Minkowski space-time, while general relativity demands a more general manifold with a metric carrying the properties of the gravitational field.  In this paper we explore the possibility of starting from a primitive notion of {\em process}, in which flux, activity or movement is taken as basic and from which physical phenomena in general and quantum phenomena in particular emerge.  Indeed we expect the space-time manifold itself to arise from this basic process.  
	
	 The problem with the traditional view is that it is mechanistic and reductionist in spirit and is in stark contrast to the perception and deep insights of Bohr \cite{nb61}, who has already argued that it is the notion of wholeness that is essential to our understanding of quantum processes, a notion that he felt could only be handled mathematically through the abstract quantum algorithm, together with the principle of complementarity.  We will argue that ideas based on a fundamental notion of process offers an alternative, more coherent view which will also provide an ontology.

The hope of finding a better understanding of Nature through a process philosophy is not new.  Already Whitehead \cite{aw} carries this analysis much further and proposes that reality is essentially an {\em organism} in which the whole determines the properties of the parts rather than the parts determining the whole.  Less well known is the work of Bohm \cite{db86} who carries these ideas much further in general terms, as well as attempting to articulate them in a mathematical form \cite{dbpdbh} \cite{bdbh84}.   Apart from the well known conceptual difficulties presented by the usual approach to quantum phenomena,  a strong motivation and support for  these ideas comes from the entangled state phenomenon known as quantum non-locality.  Here the properties of groups of individual systems cannot be derived from {\em a priori} given individual properties, but inherit properties derived from the whole system, supporting Whitehead's use of the term `ogranic'.
	
	How then are we to start to build such a theory and develop it into a sound mathematical structure?  I believe that Grassmann \cite{gg} and Hamilton \cite{wh67} had already begun to show us how this might be achieved in terms of  basic elements that form an algebra.  Indeed Grassmann was motivated by a notion of {\em becoming}, a notion that eventually led him to what we now call a Grassmann algebra.  This structure plays an important role in physics today.   However the full possibilities of Grassmann's ideas have been lost because the original motivation has been forgotten.  With this loss, the exploitation of this potentially rich structure has been stifled \cite{dh03} \cite{cdal}.

	Grassmann began his discussion by drawing a distinction between {\em form} and {\em magnitude}.  Today's physics is all about magnitude, about measured values in a given form, but for Grassmann it was about exploring and developing new forms.  The notion of form is very broad and  more commonly occurs in `thought' or  `thought form' as Grassmann put it \cite{gg3}  Now thought is about becoming and becoming yields a continuous form.  Essentially we could ask how one thought becomes another?  Is the new thought independent of the old or is there some essential dependence?  The answer to the first question is clearly ``no", because the old thought contains the potentiality of the new thought, while the new thought contains a trace of the old. Hence again as Grassmann puts it ``the continuous form is a twofold act of placement and conjunction, the two are united in a single act and thus proceed together as an indivisible unit". Why not a similar idea for the unfolding of material processes?  If we succeed we will have the possibility of removing the difference between material processes and thought.
   
	Let us try to formalise Grassmann's ideas and regard $T_{1}$ and $T_{2}$ as the opposite distinguishable poles of an indivisible process.  Notice however we have made a distinction in the overall process as Kauffman \cite{lk98} would put it.  Making a distinction does not deny the implicit indivisibility of the process, it simply notes the differences.  Therefore let us write the mathematical expression for this process as $[T_{1}T_{2}]$, the brace emphasising its indivisibility.  We can represent this brace in the form of a diagram (see Figure 1 below.)

\begin{figure}[h]
	\begin{center}
	\includegraphics[width=1.5in]{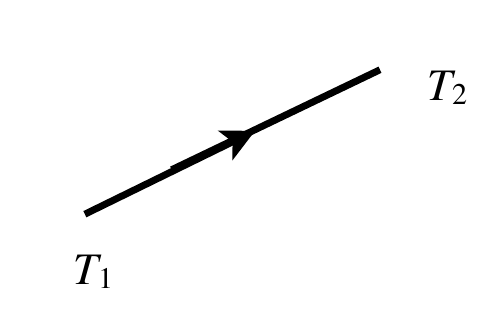}
	\caption{One-simplex.  }
	\end{center}
	\end{figure}

The arrow emphasises that  $T_{1}$ and $T_{2}$, although distinguishable, cannot be separated.  

When applied to space, we will write $[P_{1}P_{2}]$, where each  distinguishable region of space will be denoted by $P_{i}$.  Grassmann called this brace an {\em extensive}.  Thus we will call quantities like $[A,B]$ {\em extensions}.  They denote the activity of one region transforming into another.  For more complex structures, we can generalise these basic processes to those shown in Figure 2.
		
	\begin{figure}[h]
	\begin{center}
	\includegraphics[width=4in]{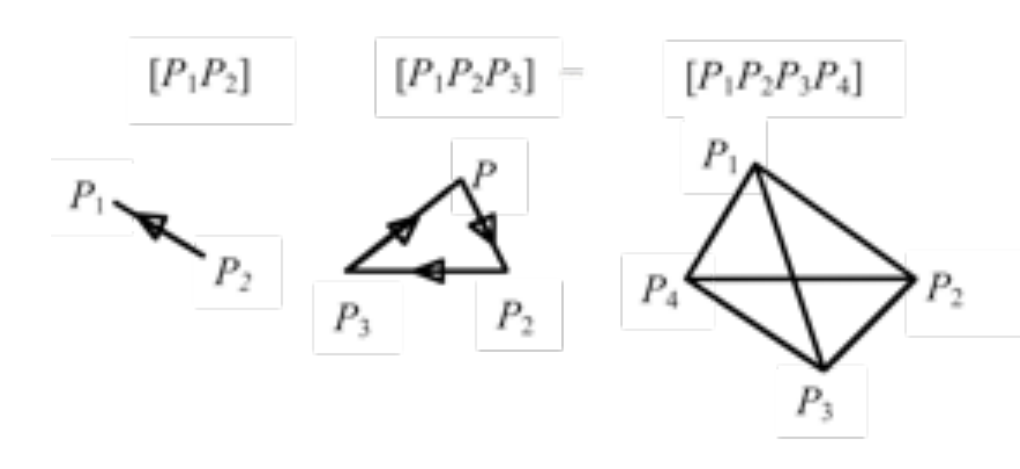}
	\caption{One, two and three-simplexes.  }
	\end{center}
	\end{figure}

In this way we have a field of extensives which can be related and organised into a multiplex of relations of  process, activity, movement\footnote{The term `movement' is being used, not to describe movements of objects but in a more general sense, implying more subtle orders of change, development and evolution of every kind\cite{db76}.}.  The sum total of all such relations constitutes what Bohm and I have called the {\em holomovement} \cite{dbbhas70}.
Thus for Grassmann, space was  a particular realisation of the general notion of process.  
	
	Let us now consider the meaning of $ [P\,P]$ where there is only one region.  This implies that region $P$ is continually transforming into itself. It is a dynamic entity essentially remaining the same, yet continually transforming into itself.  It is an invariant feature in the total flux.  Mathematically it is an idempotent as $[P\,P][P\,P]=[P\,P]$. We will see that idempotents play a central role in our approach.
	
Notice we are working with quite general notions and anything that remains invariant in a dynamical process can be treated as an idempotent.  Thus an atom is, on one level, an entity, not of `rock-like' existence, but a process continually transforming into itself.  But on another level it is composed of sub-processes, electrons, nucleons, quarks and so on, each of which can be represented by idempotents. A group of molecules that remain stable can also be treated as a relative idempotent in a more complex structure of process.  Clearly this can be extended to viruses, animals, plants right up to human beings and beyond.
	
At any level the idempotent can be thought of as a `point' which has structure, but what are the consequences if it is taken to be a point in some geometry?  Here  the notion of a point becomes a dynamic entity and not the changeless entity of classical geometry. The geometry is dynamic.  It is just what we want, for example, in general relativity, showing that there is no separation between the `geometry' and the `matter' it contains.  If we consider space to be made up of `points' then space itself cannot be a static receptacle for matter.  It is a dynamic, flowing structure of process or activity.  This is the holomovement.  In this sense it has much in common with the quantum vacuum from which quasi-invariant structures emerge and into which quasi-invariant structures decay.  
   
 Let us now introduce some mathematical structure.  Let us assume these processes form an algebra over the real field\footnote{ We assume the real field for convenience and leave open the possibility for a more fundamental structure.} in the following sense: 
   \begin{itemize}
\item multiplication by a real scalar denotes the strength of the process.  
\item the process is assumed to be oriented.  Thus  $[P_{1}P_{2}]  =  - [P_{2}P_{1}]$.  
\item next we introduce an inner multiplication defined by 
\begin{eqnarray}
[P_{1}P_{2}]\bullet
[P_{2}P_{3}]  = \pm [P_{1}P_{3}]   \label{eq:Bprod}
\end{eqnarray}
This can be regarded as the {\em order of succession}.  The choice of  $+$ or  $-$ will turn out ultimately to do with what is conventionally called the metric of the space constructed from the processes.  We will develop this idea as we go along.
\item to complete the notion of an algebra we need to define a notion of addition of two processes to produce a new process.  A mechanical analogy, although inadequate in many ways, of this is the motion that arises when two harmonic oscillations at right angles are combined.  It is well-known that these produce an elliptical motion when the phases are adjusted appropriately.  This addition process can be regarded as an expression of the {\em order of co-existence} completing our Leibnizian view of process.  We will clarify what addition means for the physics as we go along.
 \end{itemize}
 We can show that the product is associative and that the mathematical structure we have defined is a Brandt groupoid \cite{hb26}, the details of which will be discussed in subsection \ref{sec:FMS}. 
 
 We now turn our attention to showing how the symmetries of space-time are carried by the algebra and, indeed, to show how the points are ordered into an overall structure.  For the sake of simplicity we will consider how we may order a discrete structure process \cite{db65}.

\section{Some Specific Algebras of Process.}
\label{sec:SSAP}

\subsection{Quaternions.}
\label{sec:Q}

Now that we have a definition of a product to denote succession, we can sharpen up what we mean by a `point'.  Recall a point is something that continually turns into itself, so that $[P_{0}P_{0}]\bullet [P_{0}P_{0}] =[P_{0}P_{0}] $. Thus $[P_{0}P_{0}]$ represents the point $P_{0}$. Thus the `point' is an {\em idempotent} as we have discussed above.  It is interesting to note that this general notion of an {\em element of existence} was first introduced by Eddington \cite{ae58}.  

We are working in a domain of continuous process or activity.  To make a distinction we must have a form that is stable, and what makes a better stable process than that process that turns continually into itself.  Mathematically it is the idempotent, $\epsilon$, that continually turns into itself, $\epsilon\epsilon = \epsilon$ hence $\epsilon\epsilon\dots\epsilon = \epsilon$.

Now let us move on to consider two independent processes  $[P_{0} P_{1}]$  and  $[P_{0} P_{2}]$.  Suppose we want to turn $[P_{0} P_{1}]$ into $[P_{0} P_{2}]$.  How is this done?  Introduce a process $[P_{1} P_{2}]$, then using the product rule we get
\begin{eqnarray}
[P_{0} P_{1}]\bullet [P_{1} P_{2}] = [P_{0} P_{2}] \nonumber
\end{eqnarray}
Here we are using the $+$ in the product.  In subsection \ref{sec:Q} will show when it is appropriate to use the minus.
We can also show
\begin{align}
 [P_{0} P_{2}]\bullet [P_{1} P_{2}] &= - [P_{0} P_{1}]  \nonumber  \\ 
 [P_{0} P_{1}]\bullet [P_{0} P_{2}] &= - [P_{1} P_{2}]  \nonumber  \\ 
[P_{0} P_{1}] \bullet[P_{0} P_{1}] &= - [P_{0} P_{0}] = - [P_{1} P_{1}] = - 1 \nonumber \\ 
[P_{0} P_{2}]\bullet [P_{0} P_{2}] &= - [P_{0} P_{0}] = - [P_{2} P_{2}] = - 1 \nonumber
\end{align}
Note we have put $[P_{0}P_{0}]=[P_{1}P_{1}]=[P_{2}P_{2}]=1$ because in the algebra, the point behaves like the identity.  Thus there exists  three two-sided units in our algebra.  The multiplication table now closes and can be written as 
 
 \begin{eqnarray}
\begin{array}{ccccc }
    1 &\vline & [P_{0} P_{1}]& [P_{0} P_{2}]& [P_{1} P_{2}]  \\ \hline
    [P_{0} P_{1}]&\vline  &-1& -[P_{1} P_{2}]&[P_{0} P_{2}]  \\ \
    [P_{0} P_{2}]&\vline &[P_{1} P_{2}]&-1&-[P_{0} P_{1}]\\ \
    [P_{1} P_{2}]&\vline & -[P_{0} P_{2}]&[P_{0} P_{1}]&-1\\\nonumber
\end{array}
\end{eqnarray}
This structure is isomorphic to the quaternion algebra ${\cal{C}}_{0,2}$.  To show this let us make the following identification:- $[P_{0} P_{1}] =i , [P_{0} P_{2}] = j$ and $[P_{1} P_{2}] = k$ where the  $\{i,j,k\}$ are the quaternions.  The multiplication table above then becomes
\begin{eqnarray}
\begin{array}{ccccc}
    1 &\vline & i & j & k   \\ \hline
    i &\vline  &-1& -k &j  \\
    j&\vline&k&-1&-i\\
    k&\vline& -j&i&-1\\\nonumber
\end{array}
\end{eqnarray}
which should be immediately be recognised as the quaternion algebra.

Rather than use the traditional notation for the quaternions, let us use the notation that is now standard in Clifford algebra texts such as Lounesto \cite{pl97}.
Thus we will write $[P_{0} P_{1}] =e_{1} , [P_{0} P_{2}] = e_{2}$ and $[P_{1} P_{2}] = e_{21}$.  Here $ e_{21}=e_{2}e_{1}$.  We use this notation from now on.

Having established that the essentials of quaternion Clifford algebra lies at the heart of our process algebra, let us use its known properties  to generate rotations through the Clifford group.  This is achieved by inner automorphism on the elements of the algebra.  A typical element of the Clifford group is given by
\begin{eqnarray}
g(\theta)=\cos (\theta/2)+e_{12}\sin(\theta/2) \label{eq:CGP}
\end{eqnarray}
In the simple structure we are considering here, all elements of the group are of this form.   Then the rotation of any element, $A$, of the algebra is produced by
\begin{eqnarray}
A' = g(\theta)Ag^{-1}(\theta) \label{eq:IAT}
\end{eqnarray}
Suppose we choose a specific rotation with $\theta = \pi/2$, then we find
\begin{eqnarray}
g(\pi/2)=(1 +e_{12})/\sqrt{2};\qquad g^{-1}(\pi/2)=(1 -e_{12})/\sqrt{2}\nonumber
\end{eqnarray}
With these two expressions we can show exactly why the rules introduced  above actually work.  Take for example $A=e_{1}$, and put $\theta = \pi/2$, then
\begin{eqnarray}
g(\pi/2)e_{1}g^{-1}(\pi/2)=e_{1}g^{-2}(\pi/2)=-e_{1}e_{1}e_{2}=e_{2}. \nonumber
\end{eqnarray}
Translating this back into the original notation this gives
\begin{eqnarray}
[P_{0}P_{1}]\bullet[P_{1}P_{2}]=[P_{0}P_{2}] \nonumber
\end{eqnarray}
which is where we started.

Clearly what we have constructed is a method of rotating through right angles in a plane.  However if we allow $\theta$ to take continuous values $0\geq \theta \geq 2\pi$, we can create a continuum of points on a circle, since
\begin{eqnarray}
e_{1}' = g(\theta)e_{1}g^{-1}(\theta)=\cos^{2}(\theta/2)-\sin^{2}(\theta/2)e_{1} + 2\cos(\theta/2)\sin(\theta/2)e_{2}\nonumber 
\end{eqnarray}
or
\begin{eqnarray}
e_{1}'=\cos(\theta)e_{1} + \sin(\theta)e_{2}\nonumber
\end{eqnarray}
Of course this relation looks very familiar but notice we are approaching things from a different point of view.  We can formalise this by  introducing a mapping from our algebra of process, our Clifford algebra, ${\cal{C}}_{0,2}$ to a Euclidean vector space $V$, viz:
\begin{eqnarray}
\eta : {\cal{C}} \rightarrow V \qquad \mbox{such that}\quad e_{1}=\hat{e}_{1}\;\mbox{and}\;e_{2}=\hat{e}_{2} \nonumber
\end{eqnarray}
where $\{\hat{e}_{1},\hat{e}_{2}\}$ is a pair of othonormal basis vectors in $V$ satisfying $\hat{e}_{i}^{2}=1; (\hat{e}_{1}.\hat{e}_{2})=0$. The inner automorphism in ${\cal{C}}_{0,2}$ then induce rotations in $V$.  Thus we have reversed the traditional argument of starting from a given vector space and constructing an algebra.  We claim that it is also possible and, for our purposes, profitable to construct a vector space from the group of of movements, or processes.  This is a more primitive example of the Gel'fand construction \cite{jdmh}.  We suggest that this possibility would be worth exploring when investigating so called quantum space-times.

So far we have confined our attention to a very simple case where we only have two degrees of freedom.  In section \ref{sec:PC} we will show how the above rules  can be generalised and applied to three basic movements  $[P_{0} P_{1}]  , [P_{0} P_{2}]$ and $[P_{0}P_{3}]$ to produce the Pauli Clifford, which in turn produces a 3-dimensional vector space, $R_{3,0}$.  The processes here are all of a polar kind.  

Let us now generalise our approach by including processes of a temporal kind, so that we are led to a Lorentz-type Clifford groups.  In the next sub-section we show exactly how this can be done, again by simply considering two basic processes $[P_{0}P]$ and $[P_{0}T]$, the latter being a temporal process. 


\subsection {Lorentz Group.}
\label{sec:LG}
Let us illustrate how to generalise our approach by considering one movement in space and one in time.  We will show that this results in the Clifford algebra ${\cal{C}}_{1,1}$ which will enable us to construct the the mini-Minkowski space $R_{1,1}$. 

In this case the basic processes are $[P_{0} P]$ and $[P_{0} T]$.  We will follow Kauffman \cite{lk98} and call $[P_{0} P]$ a {\em polar} extensive while $[P_{0} T]$ will be called a {\em temporal} extensive.  We now extend the product to
\begin{align}
[P_{0} P] \bullet[P_{0} P] &= - [P_{0} P]\bullet [P P_{0}] = - [P_{0} P_{0}] = - 1 
\nonumber\\ \
[P_{0} T]\bullet [P_{0} T] &= - [P_{0} T]\bullet [T P_{0}] = + [P_{0} P_{0}] = + 1
\nonumber
\end{align}
Here we have used the relations $ [P_{0} P_{0}] =1;\;[P\;P] = 1$ while $[T\;T]=-1$.  The difference in sign between these two terms will ultimately emerge in the signature of the metric $g_{ij}$ when constructed.  Thus we have $[P\,P]=g_{PP}$ and $[T\,T]=g_{TT}$.

We then get the multiplication table
 \begin{eqnarray}
\begin{array}{ccccc }
    1 &\vline & [P_{0} T]& [P_{0} P]& [P\: T]  \\ \hline
    [P_{0} T]&\vline  &1& [P\: T]&[P_{0} P]  \\ \
    [P_{0} P]&\vline &-[P\: T]&-1&[P_{0} T]\\ \
    [P\: T]&\vline & -[P_{0} P]&-[P_{0} T]&1\\\nonumber
\end{array}
\end{eqnarray}
 If we identify $[P_{0} T] = e_{0}, [P_{0} P]=e_{1}$ and $[P\, T] =e_{10}$,  we see this multiplication table is isomorphic to the multiplication table of the Clifford algebra ${\cal{C}}_{1,1}$.
 
 In this algebra it is of interest to examine the sum, $[P_{0}T] + [P_{0}P]$, and the difference  $[P_{0}T] - [P_{0}P]$.  Here
 \begin{eqnarray}
\Bigl[[P_{0}T] \pm [P_{0}P]\Bigr]\bullet [P\,T]=-[P_{0}T]\bullet[T\,P]\pm[P_{0}P]\bullet[P\,T]=\Bigl[[P_{0}T] \pm [P_{0}P]\Bigr] \nonumber
\end{eqnarray}
In other words $\Bigl[[P_{0}T] \pm [P_{0}P]\Bigr]$ transforms into itself. Now writing this in a more familiar form, we have
\begin{eqnarray*}
\Bigl[[P_{0}T] \pm [P_{0}P]\Bigr] =e_{0} \pm e_{1}\quad \xrightarrow{\eta}\quad t \pm x 
\end{eqnarray*}
which we should immediately recognise as the light cone co-ordinates used by Kauffman \cite{lk98}.  In our approach $\Bigl[[P_{0}T] + [P_{0}P]\Bigr]$ corresponds to a movement along one light ray, while $\Bigl[[P_{0}T] - [P_{0}P]\Bigr]$ corresponds to a movement along a perpendicular light ray as shown in figure 3.

	\begin{figure}[h]
	\begin{center}
	\includegraphics[width=4in]{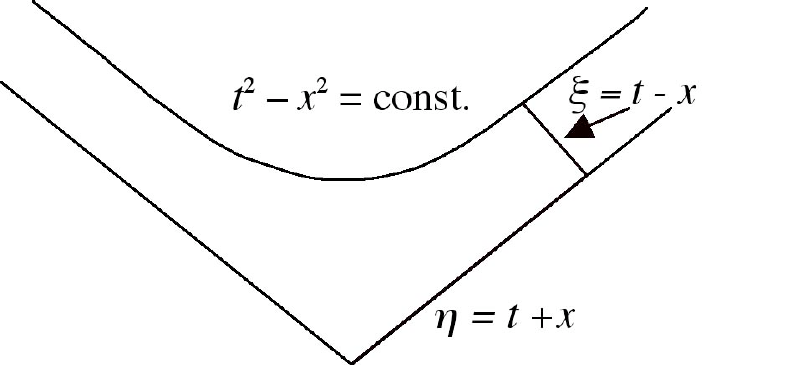}
	\caption{Light Cone Coordinates.  }
	\end{center}
	\end{figure}
	
 
Since we have a movement in time and a movement in space, how then do we envisage the notion of velocity?  In other words can we give a meaning to dividing one vector by another?  Suppose we first form the reciprocal  $1/ [P_{0}T]$.  Then clearly
\begin{eqnarray*}
[P_{0}T]^{-1}=\frac{1}{[P_{0}T]}\times\frac{[P_{0}T]}{[P_{0}T]}=\frac{[P_{0}T]}{[P_{0}T]}\times\frac{1}{[P_{0}T]}=[P_{0}T]
\end{eqnarray*}
 But leaves us with a problem.  Do we write for the velocity
\begin{eqnarray*}
 [P_{0}P]\bullet [P_{0}T]^{-1}=-[P\,T] \qquad \mbox{or} \qquad  [P_{0}T]^{-1}\bullet[P_{0}P]= [P\,T]?
\end{eqnarray*}
The conventional approach through the Clifford algebra identifies the positive direction of the velocity with $\alpha_{1}=\gamma_{01}$ in the case of the Dirac theory.  In our simple Clifford algebra ${\cal{C}}_{1,1}$, the equivalent to $\alpha_{1}$ is $ e_{01}=[P_{0}P]\bullet[P_{0}T]^{-1}$.  We then identify $e_{01}$ with the  positive direction of the velocity\footnote{Note this immediately gives us an explanation as to  {\em why} we identify $\alpha_{0i}$ with the velocity in the Dirac theory.}.  With this identification we can write an element of the Clifford group as
 \begin{eqnarray*}
g^{\pm}(\lambda) = \cosh(\lambda/2) \pm e_{01}\sinh(\lambda/2) 
\end{eqnarray*}
where $\lambda$, the rapidity is given by $\tanh(\lambda)=v/c$.  In terms of these transformations it is easily shown that
\begin{eqnarray*}
e'_{0} +e'_{1} = g(e_{0} +e_{1}) g^{-1} =k^{-1}(e_{0} +e_{1}) 
\end{eqnarray*}
and
\begin{eqnarray*}
e'_{0} - e'_{1} = g(e_{0} - e_{1}) g^{-1} =k(e_{0} - e_{1}) 
\end{eqnarray*}
where $k= \sqrt{\frac{1+v}{1-v}}$.  Since we can write 
\begin{eqnarray*}
e_{0}\xrightarrow{\eta}t \qquad \mbox{and}\qquad e_{1} \xrightarrow{\eta} x
\end{eqnarray*}
we see that
\begin{eqnarray*}
t'+x' =k^{-1}(t + x)\qquad \mbox{and} \qquad t' - x' = k(t-x)
\end{eqnarray*}
This is just the defining expression for the $k$-calculus from which Kauffman \cite{lk98} extracts all the transformations required for special relativity.  This result also holds when ${\cal{C}}_{1,1}$ is generalised to ${\cal{C}}_{1,3}$.  Thus we provide an alternative but complementary approach to the Kauffman approach.  We will discuss Kauffman's approach to special relativity a little further later in the paper.  


\subsection{The Pauli Clifford}
\label{sec:PC}

Let us now move onto generating the Pauli Clifford.  As we have already remarked  we need 
	three basic polar extensives  $[P_{0}P_{1}], [P_{0}P_{2}], [P_{0}P_{3}]$.  We require these to describe three independent movements.  We also require three movements, one to take us from   $[P_{0}P_{1}]$ to  $[P_{0}P_{2}]$, another to take form  $[P_{0}P_{1}]$ to $[P_{0}P_{3}]$ and a third to take us from  $[P_{0}P_{2}]$ to $[P_{0}P_{3}]$. Call these movements $[P_{1}P_{2}], [P_{1}P_{3}]$ and $[P_{2}P_{3}]$.  
	
	If we are to construct the Pauli algebra we need to assume the following relations,
\begin{eqnarray}
[P_{0}P_{0} ]=1; \quad[P_{1}P_{1}]= [P_{2}P_{2}] = [P_{3}P_{3}] =-1 \nonumber
\end{eqnarray}
We should notice the change in sign in this case.  This is to take account of the different metric we are choosing and because of that we need to ensure $[P_{0}P_{1}]\bullet[P_{0}P_{1}] =[P_{0}P_{2}]\bullet[P_{0}P_{2}] =[P_{0}P_{3}]\bullet[P_{0}P_{3}] = 1$.
At this stage the notation is looking a bit clumsy so it will be simplified by writing the six basic  movements as $[a], [b], [c], [ab], [ac], [bc]$.
 
	We now use the order of succession to establish
 \begin{eqnarray}
 \nonumber  	 [aa] = [bb] = [cc] = 1\hspace{0.6cm}\\
 \nonumber	 [ab]\bullet[bc] =[ac] \hspace{1.2cm}  \\ 
 \nonumber	 [ac]\bullet[cb] = [ab]  \hspace{1.2cm} \\
 \nonumber	 [ba]\bullet[ac] = [bc]  \hspace{1.2cm} 
  \end{eqnarray}
where the rule for the product is self evident. There exist in the algebra, three two-sided units, $[aa], [bb]$, and $[cc]$.  For simplicity we will replace these elements by the unit element 1.  This can be justified by the following results  $[aa]\bullet[ab] = [ab]$;  $[ba]\bullet[aa] = [ba];   [bb]\bullet[ba] = [ba]$, etc.

Now 
\begin{align}
\nonumber	         [ab]\bullet[ab] &= - [ab]\bullet[ba] = [aa] =  -1\\
\nonumber		[ac]\bullet[ac] &= - [ac]\bullet[ca] = [cc] =  -1 \\
\nonumber		[bc]\bullet[bc] &= - [bc]\bullet[cb] = [bb] = - 1
\end{align} 
There is the possibility of forming $[abc]$.  This gives
\begin{align}
\nonumber [abc]\bullet[abc] &= - [abc]\bullet[acb] = [abc]\bullet[cab] =- [ab]\bullet[ab] = - 1\\
\nonumber		[abc]\bullet[cb] &= -[ab]\bullet[b] = [a], \qquad \mbox{etc}.
\end{align}
In the last expression we need to know that $[P_{0}P_{i}]\bullet[P_{j}P_{k}] = [P_{j}P_{k}]\bullet[P_{0}P_{i}] \quad \forall \;i \ne j \ne k. $
Thus the algebra closes on itself.  If we now change the notation to bring it in line with standard Clifford variables we have 
\begin{eqnarray}
[P_{0}P_{i}] = e_{i}; \quad [P_{i}P_{j}] = e_{ji}; \quad [abc] = -e_{123} \nonumber
\end{eqnarray}
then it is straight forward to show that the algebra is isomorphic to the Clifford algebra $ R_{3,0}$ which we was called the Pauli-Clifford algebra in Frescura and Hiley (1980).

	The significance of this algebra is that it carries the rotational symmetries of a three-dimensional Euclidean space. The movements $[ab], [ac]$ and $[bc]$ generate the Lie algebra of SO(3), the group of rotations in three-space, but because we have constructed a Clifford algebra, the rotations are carried by inner automorphisms of the type shown in equation (\ref{eq:IAT}).  A typical form of $g(\theta)$ is shown in equation (\ref{eq:CGP}).  In the three dimensional case we use the bivectors $[ab] = e_{21}, [ac] = e_{31}$ and $[bc] = e_{32}$ to define three infinitesimal rotations about the three independent axes.
	
	We can extend the method we have outlined above to generate the Dirac Clifford algebra and the conformal Clifford algebra which contains the twistor originally introduced by Penrose \cite{rp67}.  We will not discuss these structures in this paper.
	
In summary then we have shown, or rather sketched out how to generate the Clifford algebras from a primitive notion of process or activity. A more rigorous mathematical approach to these topics will be found in Griffor \cite{ag00}.  The algebras we have discussed in this paper so far only use orthogonal groups and so we have, in effect, developed 
a {\em directional calculus}.  What is missing from our discussion so far is the translation symmetry.
The generalisation to include translations will be discussed later


\section{Connections with Other Mathematical Approaches.}

\subsection{Formal Mathematical Structure}
\label{sec:FMS}

We have already indicated that we have identified the structure we are exploring as a Brandt groupoid \cite{hb26} but we don't have to return to this early work to get a formal understanding of this structure.  Ronnie Brown \cite{rb87} has an excellent review of this structure which we will briefly explain here.

A groupoid $G$ is a small category in which every morphism is an isomorphism.  This category comprises a set of morphisms, together with a set of objects or points, $Ob(G)$. We also have a pair of functions 
$s, t : G \rightarrow Ob(G),$ together with a function $ i : Ob(G) \rightarrow G$ such that $si = ti = 1$. The functions $s, t$ are called
the source and target maps respectively. 

The elements of process that we have been using can easily be connected with this language.  Consider a general movement $a_{j}$.  We can identify $sa_{j}$ as $P_{0}$ and $ta_{j}$ as $P_{j}$, then 
\begin{eqnarray*}
[P_{0}P_{j}]\qquad \Rightarrow \qquad a_{j} : sa_{j} \rightarrow ta_{j}
\end{eqnarray*}
If we have a pair of movements $a$ and $b$ then we can compose these movements if $ta =sb$. Thus
\begin{eqnarray*}
[sa \xrightarrow{a} ta][sb \xrightarrow{b} tb] = [sa \xrightarrow{ab} tb]\qquad \mbox{iff}\quad ta = sb.
\end{eqnarray*}
It is not difficult to show that this product is associative. Thus the mathematical structure we  considered in the last section can be formally identified with a groupoid.

It should be noted that our approach is also related to the approach of Abramski and Coecke \cite{sabc04} who apply category theory to quantum mechanics, again attempting to build a process view of quantum phenomena.  It would be interesting to compare the two approaches in detail but such a comparison will not be attempted in this paper. 

\subsection{Comparison with Kauffman's Calculus of Distinctions.}

In the earlier sections of this paper, we have referred to the work of Kauffman \cite{lk98} \cite{lk87} \cite{lk82} which provides us with a calculus of distinctions, or as it is sometimes called the iterant algebra.  Kauffman also introduces a binary symbol $[A,B]$ which he motivates in the following way.  He wants to start with a basic idea from which all further discussions follow.  Start with an undifferentiated two-dimensional canvas.  Then make a distinction by defining a boundary, dividing the canvas into two regions marking  the `inside' $A$ and  the `outside' $B$ symbolizing  this distinction by $[A, B]$. 

How is this distinction related to the basic movement  that we have used earlier? 
 To bring out this similarity, we can also write this as  $\UR{A} = B$. Here we use $\UR{X}$ to denote Spencer Brown's `cross' indicating we must cross the boundary.   Kauffman \cite{lk98} uses this analogy to argue that not only do we make a distinction but we must act out the distinction by actually crossing the boundary.  Crossing the boundary enables us to discuss change, to discuss movement, to discuss process. 

Now in order to structure these movements, Kauffman introduces a multiplication rule
\begin{eqnarray}
[A,B]*[C,D] \coloneq [AC,BD] \label{eq:Kprod}
\end{eqnarray}
Furthermore he assumes that $AC =CA$ and $BD =DB$, i.e. they are commutative products.  In this case, suppose we take $B=C$ then
\begin{eqnarray*}
[A,B]*[B,C]=[AB,BC]=B[A,C]
\end{eqnarray*}
Then if we put $B = \pm 1$ we obtain the same product rule that we introduced earlier in this paper.  Thus we see that our algebra is a special case of that introduced by Kauffman.

The sum that Kauffman introduces is also a generalisation of the sum we introduce.  He defines
\begin{eqnarray*}
[A,B]+[C,D]\coloneq[A+C,B+D]
\end{eqnarray*}
In Kauffman \cite{lk87} the $k$-calculus is set up by directly considering $[t +x]$ and $[t-x]$ but it is difficult to understand how `marks' $A$ and $B$ become co-ordinates.  In section \ref{sec:Q} above we have a clearer way of showing how the co-ordinates arise since we have a well defined mapping, $\eta$, taking us from the Clifford algebra, ${\cal{C}}_{n,m}$, to an underlying vector space
$V_{n,m}$.  Once again notice the order, ${\cal{C}}_{n,m}$ is primary, $V_{n,m}$ is derived.

\subsection{Some Deeper Relations between our Approach and that of the Calculus of Distinction.}
\label{sec:SDR}

Let us go deeper into the relation between the general structure of Clifford algebras and its relationship to the Kauffman calculus \cite{lk98}.
However we first want to point out that Sch\"{o}nberg \cite{ms57} and Fernandes \cite{mf95} have shown us how to build any
orthogonal Clifford algebra from a pair of dual Grassmann algebras whose
generators satisfy the relationship
\begin{eqnarray}
[a_{i},a^{\dag}_{j}]_{+} =g_{ij}\hspace{1cm}[a_{i},a_{j}]_{+} =0=[a^{\dag}_{i},a^{\dag}_{j}]_{+} 
\end{eqnarray}

These will be recognised as {\em vector} fermionic `annihilation' and `creation'
operators\footnote{This suggests that these operators could be used to describe the creation and annihilations of extensions.}. Notice these are vector operators and not the spinor operators
used in particle physics. Using these we find
\begin{eqnarray}
\sigma_{x}= a + a^{\dag}\hspace{0.7cm} \mbox{and}\hspace{0.7cm} \sigma_{z} = a - a^{\dag}
\end{eqnarray}
 
 Now Kauffman \cite{lk98} introduces an operation  $p$ that destroys the inside but leaves the outside alone.  This can be written as
\begin{eqnarray}
p\ast[A,B] = [0,B]. \label{eq:ast}
\end{eqnarray}
One way to do this is to let $p = [0,1]$ and use the product $\ast$ defined in equation \eqref{eq:Kprod}.
Then clearly equation (\ref{eq:ast}) will follow.  Kauffman develops these ideas and shows 
\begin{eqnarray}
\sigma_{x}\ast[A,B] = [B,A] \hspace{1cm}\mbox{and}\hspace{1cm}\sigma_{z}\ast[A,B]=[A,-B]
\end{eqnarray}
Now let us ask what our operators $a$ and $a^{\dag}$ produce when we operate on $[A,B]$. It is straightforward to show
\begin{eqnarray}
a\ast [A,B] = [B,0]\hspace{0.7cm}\mbox{and}\hspace{0.7cm}
a^{\dag}\ast[A,B] = [0,A]
\end{eqnarray}
Thus we see that here the annihilation operator $a$ destroys the inside and puts
the outside inside. On the other hand the creation operator $a^{\dag}$ destroys the outside and puts the inside outside!

We can actually carry this further and ask what action the algebraic spinors
(minimal left ideals) have on $[A,B]$. It is not difficult to show that in this case
we have two algebraic spinors given by
\begin{eqnarray}
\psi_{L_{1}} = aa^{\dag} + a^{\dag} \hspace{0.7cm}\mbox{ and}\hspace{0.7cm} \psi_{L_{2}} =  a^{\dag}a + a.
\end{eqnarray} 
Now let us ask what happens when we use these spinors to  produce a change
in $ [A, B]$. What is this change? Again the answer is easy because
\begin{eqnarray}
\psi_{L_{1}}\ast[A, B] = (aa^{\dag}+a^{\dag})*[A,B] = [A, A] \\
\psi_{L_{2}}\ast[A, B] = (a^{\dag}a+a)\ast[A, B] = [B, B] 
\end{eqnarray}
This means that one type of spinor, $\psi_{L_{1}}$, destroys the outside and puts a copy
of the inside outside, while the other spinor, $\psi_{L_{2}}$, destroys the inside and puts
a copy of the outside inside! In other words these operators remove the
original distinction, but in different ways. More importantly from the point
of view that we are exploring here is that we see how the algebraic spinor
itself is active in producing a specific change in the overall process.

Physicists treat these two spinors, $\Psi_{L_{1}}$ and $\Psi_{L_{2}}$ as equivalent and map them onto an external spinor space, the Hilbert space of ordinary spinors $\psi$. In other words this single spinor, $\psi$, is an equivalence class of
the algebraic spinor when it is projected onto a Hilbert space (see Bratteli
and Robinson \cite{br79}). The projection means that we have lost the possibility
of exploiting the additional dynamical structure offered by the algebraic spinors.

Notice that these spinors are themselves part of the algebra. The whole thrust
of our argument is that we must exploit the properties of the algebra and not
confine ourselves to an external Hilbert space. When we do this we can continue with
our idea that the elements of our algebra describe activity or process even in the quantum domain.

Let us take this whole discussion a little further. Consider the following relationship
\begin{eqnarray}
a\ast[A, 0] = [0, 0] 
\end{eqnarray}
This should be compared with the physicist's definition of a vacuum state,
\begin{eqnarray}
a|0\rangle = a\Downarrow = 0 \label{eq:vac}
\end{eqnarray}
where we have introduced the notation used by Finkelstein \cite{df96}. Thus
equation (\ref{eq:vac}) shows that $[A, 0]$ acts like the vacuum state in physics.

Furthermore $
a^{\dag}\ast[0,B] = [0,0] $
should be compared with $a^{\dag}\Uparrow = 0$.
Finkelstein calls, $\Uparrow$, the plenum. Thus we see that $[0, B]$ acts like the
plenum. Thus we see that here the vacuum state is not empty. Internally it
has content but externally it is empty. For the plenum, it is the other way
round, so unlike Parmenides we can have `movement' from outside to inside!
We can take this a bit further by recalling that we can write the projector
onto the vacuum as $V = |0\rangle\langle0|$, then we have $aV = 0$. If the projector onto
the plenum is $P = |\infty\rangle\langle\infty|$, we find $a^{\dag}P = 0$.  It is interesting to note that these relations are central to the approach of Sch\"{o}nberg \cite{ms57}.

\subsection{Extensions and the Incident Algebra of Raptis and Zapatrin.}

All of the above discussion suggests that we could write $[A , B ]$ as $|A\rangle\langle B|$. It should be noticed that this is only a symbolic change and the bra/ket should not be identified with vectors in an external vector space.  Nevertheless by making this
identification we can bring out the relationship of our work to that of
Zapatrin \cite{rz01} and of Raptis and Zapatrin \cite{jrz} who developed an approach through what they called the incident algebra. In this structure the product rule is written in the form
\begin{eqnarray*}
|A\rangle\langle B|\cdot|C\rangle\langle D|=|A\rangle\langle B|C\rangle\langle D|=\delta_{BC}|A\rangle\langle D|
\end{eqnarray*}

Again this multiplication rule is essentially rule \eqref{eq:Bprod}, the order of succession above. But there is a major difference. When $B \ne C$ the product equals zero, whereas we leave it undefined at this level. So tempting as it seems we must not {\em identify} their $|A\rangle\langle B|$ with our use of the same symbols.  Nevertheless a good notation can take us to places where we did not expect to go as we will see!

\section{Some Radical New Ideas.}

\subsection{Intersection of the Past with the Future.}

We now want to look more deeply into the structure based on relationships like
$[P_{1} P_{2}]$ by, as it were, `getting inside' the connection between $P_{1}$ and $P_{2}$.
Remember we are focusing on process or movement and we are symbolising the notion of {\em becoming}
by $[P_{1} P_{2}]$. Ultimately we want to think of these relationships as an ordered
structure defining what we have previously called `pre-space' (see Bohm \cite{db86} and Hiley \cite{bh91}). In other words, these relationships are not to be
thought of as occurring in space-time, but rather the properties of space-time are to be abstracted from this pre-space.  We have already suggested how we can achieve this through the mapping $\eta : {\cal{C}} \rightarrow V$ but much more work has to be done.  Let us go more slowly.

Conventionally physical processes are always assumed to unfold in space-time, and furthermore,  time evolution is always assumed to be from point to point. In other words, physics always tries to talk about time-development {\em at an instant}. Any change always involves the limiting process $\lim_{\Delta t \rightarrow 0}(\Delta x)/(\Delta t)$. 

 However before taking the limit we were taking points in the past $(x_{1}, t_{1})$ and relating them to a points in the future $(x_{2}, t_{2})$, that is we are relating what {\em was} to what {\em will be}. But we try to hide the significance of that by going to the limit  $t_{2} - t_{1} \rightarrow 0$ when we interpret the change to take place at an instant, $t$. Yet curiously the instant is a set of measure zero sandwiched between the infinity of that which has passed and the infinity of that which is not yet.  This is fine for evolution of point-like entities but not for the evolution of structures.

At this point I wish to recall Feynman's classic paper where he sets out his thinking that led to his `sum over paths' approach \cite{rf}. There he starts by dividing space-time into two regions $R'$ and $R^{\prime\prime}$. $R'$ consists of a region of space occupied by the wave function before time $t'$, while $R^{\prime\prime}$ is the region occupied by the wave function after time $t^{\prime\prime}$, i.e $t' < t^{\prime\prime}$. Then he suggested that we should regard the wave function in region $R'$ as contain information coming from the `past', while the conjugate wave function in the region $R^{\prime\prime}$ represents information coming from the `future'\footnote{This is essentially the same idea that led to the notion of the anti-particle `going backward in time', but at this stage we are not considering  anti-matter.}. The possible `present' is then the intersection between the
two, which is simply represented by the transition probability amplitude
$\langle\psi (R^{\prime\prime})| \psi(R')\rangle $. But what I want to discuss here is $|\psi (R')\rangle\langle \psi (R^{\prime\prime})|$. This is
where all the action is!

Before taking up this point I would like to call attention to a similar notion
introduced by Stuart Kauffman \cite{sk00} in his discussion of biological
evolution.  Here the discussion is about the evolution of {\em structure}. He talks about the evolution of
biological structures from their present form into the `adjacent possible'.
This means that only certain forms can develop out of the past. Thus not
only does the future form contain a trace of the past, but it is also
constrained by what is `immediately' possible, its {\em potentialities}. So any development is
governed by the {\em tension between the persistence of the past, and an anticipation of the future}.

What I would now like to do is to build this notion into a dynamics.
Somehow we have to relate the past to the possible futures, not in a completely deterministic way, but in a way that constrains the possible future development. My basic notion is that physical processes involve an extended structure in both space and time. I have elsewhere called this structure a `moment' \cite{bh01}. In spatial terms, it is fundamentally non-local in space, however it is also `non-local' in time. I see this more in terms of  a-local concepts, with locality yet to be defined. It is a kind of `extension in time' or a `duron', an idea that has a resonance with what Grassmann was trying to do as we outlined earlier. This extension in time may seem too extreme but remember quantum theory must accommodate the energy-time uncertainty principle. This implies that a process with a given sharp energy cannot be described as unfolding at an instant of time except, perhaps, in some approximation.  The notion of a moment captures the essential ambiguity implicit in quantum theory.

The idea of process that we introduced in section \ref{sec:SSAP} is naturally suited to describing this notion of a moment.  All we need to do is to regard the two aspects of $[P_{1}P_{2}]$ as functions of two times so that we can write this bracket in the form $[A(t_{1}),B(t_{2})]$ and show that, as Feynman \cite{rf} actually demonstrates for the Schr\"{o}dinger case, we capture the usual equations of motion in the limit $t_{1}\rightarrow t_{2}$.

\subsection{A Change in Notation.}

In section \ref{sec:SDR} we have already shown that if write $|0\rangle\langle0|$ for $[0,0]$ we have the possibility of a ready made link between our notation and the bra-ket notation of Dirac.  In fact Dirac in his work on spinors \cite{pd74} has already suggested that we can factorise an element of the Clifford algebra into a pair $|A\rangle$ and $\langle B|$.  He shows that this ket is, in fact, a spinor, while the bra is a dual spinor.

Let us therefore follow Dirac and replace $[P_{1},P_{2}]$  by $|P_{1}\rangle\langle P_{2}|$\footnote{ $|P_{1}\rangle$ and $\langle P_{2}|$ are not to be taken as elements in a Hilbert space and its dual.}.  In section \ref{sec:Q} we introduced the Clifford group without any justification in terms of the ideas we were developing.  Rather we called on our prior knowledge of the structure of Clifford algebras to introduce the idea.  However we can justify choice once we have assumed that the basic process can be factorised.  Now we can argue that under a rotation, we have the transformations
\begin{eqnarray*}
|P'\rangle = R |P\rangle \qquad \mbox{and}\qquad \langle P'| = \langle P| R^{-1}
\end{eqnarray*}
This gives us the added advantage of showing immediately that $\langle P|P \rangle$ is  rotationally invariant.

This also helps us the understand Kauffman's \cite{lk87} \cite{lk01} notation when he writes the Lorentz transformation in the form
\begin{eqnarray*}
{\cal{L}}[A,B]= [k^{-1}A,kB]
\end{eqnarray*}
where $k=\sqrt{\frac{1+v}{1-v}}$ is a function of the relative velocity.  First we consider a boost in the $x$-direction, $g(v)$, and write
\begin{eqnarray}
{\cal{L}}[A,B] = g(v)|A\rangle\langle B|g^{-1}(v)
\end{eqnarray}
which does not in general reduce to Kauffman's expression.  However if we form
\begin{eqnarray*}
e'_{0}+e'_{1} = g(v)\Bigl(|P_{0}\rangle(\langle T|+\langle P|)\Bigr)g^{-1}(v) =k^{-1}\Bigl(|P_{0}\rangle(\langle T| +\langle P|)\Bigr) = k^{-1}(e_{0}+e_{1})
\end{eqnarray*}
and similarly
\begin{eqnarray*}
e'_{0}-e'_{1} = g(v)\Bigl(|P_{0}\rangle(\langle T|-\langle P|)\Bigr)g^{-1}(v) =k\Bigl(|P_{0}\rangle(\langle T| -\langle P|)\Bigr) = k(e_{0}-e_{1})
\end{eqnarray*}
We arrive at Kauffman's result provided we write $A = |P_{0}\rangle \langle T| +|P_{0}\rangle \langle P|$  and  $B=|P_{0}\rangle \langle T| -|P_{0}\rangle \langle P|$.

The traditional way to proceed is to assume that $|P\rangle $ is a vector in an external Hilbert space.  As Frescura and Hiley \cite{ffbh80} have shown there is, in fact, no need to go outside the algebra.  We can identify the ket as an element, $\Psi_{L}$, of a left ideal, ${\cal{I}}_{L}$, in the algebra.  One way to generate such an element is to choose a suitable primitive idempotent, $\epsilon$.  Then we multiply from the left by any element of the algebra to form the element $\Psi_{L}$ giving
\begin{eqnarray*}
\Psi_{L} =A \epsilon
\end{eqnarray*}
where the general element $A$ can be written in the form  $A = a_{0} +\sum a_{i}e_{i} + \sum a_{ij}e_{ij} + \dots + a_{n}e_{12\dots n}$, when the Clifford algebra is generated by $\{1, e_{1}\dots e_{n}\}$\footnote{Physicists should not be put off by the notation because the $e$s are exactly the $\gamma$s used in standard Dirac theory  We have changed notation simply to bring out the fact that we do not need to go to a matrix representation, although that possibility is always open to us should we wish to take advantage of it}.  There is a simplification with which we will avail ourselves and that is that, in the Clifford algebras in which we are interested,  we can generate an element of the left ideal by choosing $A $ to comprise only the even elements of the algebra.  In this case we can write   
\begin{eqnarray*}
A = \psi_{L} = a_{0} + \sum a_{ij}e_{ij} + \sum a_{ijkn}e_{ijkn} + \dots .
\end{eqnarray*}

The corresponding dual element corresponding to $\langle B|$ is an element of the minimal right ideal generated from $\epsilon$ and is given by
\begin{eqnarray*}
\Psi_{R}=\epsilon B
\end{eqnarray*}
We can also obtain $\Psi_{R}$ from the  element of the minimal left ideal by {\em conjugation}.  Conjugation is defined as an anti-involution on $\cal{A}$ induced by the orthogonal involution $-1_{X}$, $X$ being the vector space induced by the Clifford algebra. In this case  $\Psi_{R}=\widetilde{\Psi}_{L}$.   Thus in our new notation, we have
\begin{eqnarray*}
[A,B] \rightarrow \Psi_{L}\Psi_{R}= \psi_{L}\epsilon\psi_{R}=\hat{\rho}
\end{eqnarray*}
We will call this element the {\em Clifford density element}.  $\hat{\rho}$ is a key element in our Clifford algebra approach to quantum theory \cite{bh08a} \cite{bhrc08a} \cite{bhrc08b}.  As we showed in those papers, we can reproduce all the results of quantum mechanics without the need to introduce Hilbert space and the wave function.  Furthermore as we will show later, we are led directly to the Bohm model showing that this model is merely a re-formulation of quantum theory in which the complex numbers are not necessary.  Before embarking on this discussion we will continue to explore the structure of space-time offered by our methods. 

\subsection{The Lorentz transformations from light rays and a clock.}

In this sub-section we  sketch how Kauffman's use of the $k$-calculus, introduced originally by Page \cite{page} and popularised by Bondi \cite{hb}, emerges from the results obtained in sub-section \ref{sec:LG} and in the previous sub-section. We have shown that the movements along light rays lead directly to the light cone co-ordinates.  We did this by noting that the light cone movements $e_{0}\pm e_{1}$ transform through
\begin{eqnarray*}
e'_{0} \pm e'_{1}= k^{\mp}(v)(e_{0} \pm e_{1})
\end{eqnarray*}
where $k(v) =\sqrt{\frac{1+v}{1-v}}$.  We then use the projection $\eta$ to obtain the light cone co-ordinates so that
\begin{eqnarray*}
t' \pm x' =k^{\mp}(v)(t \pm x)
					\label{eq:k}
\end{eqnarray*}

The basic idea of the $k$-calculus  is to explore the geometry of space armed only with a radar gun and a clock.  The gun is fired from the origin of the co-ordinate system at time $t_{1}$ so that this event has co-ordinates $(t_{1}, 0)$.  The radar signal returns to the origin at time $t_{2}$ after being reflected off an object at the point $(t, x)$.  Assuming the speed of light is unity $(c = 1)$ both ways, we see that
\begin{eqnarray*}
t_{2} - t_{1} = 2x\qquad\mbox{and}\qquad t_{2} +t_{1}=2t.
\end{eqnarray*}
In terms of the Kauffman's iterant calculus, we can write 
\begin{eqnarray*}
[t_{2}, t_{1}]=[t+x, t-x]
\end{eqnarray*}
But we have already shown in equation \eqref{eq:k} how the RHS of this equation transforms under a Lorentz transformation.   Thus we have
\begin{eqnarray*}
[t^{\prime}_{2}, t'_{1}]=[t'+x', t'-x']=[k^{-1}(v)(t+x),k(v)(t-x)] = [k^{-1}(v)t_{2}, k(v)t_{1}]
\end{eqnarray*}
This means that
\begin{eqnarray}
t'_{1}=kt_{1}(v)\qquad\mbox{and}\qquad t_{2} =k(v) t'_{2}
				\label{eq:tk}
\end{eqnarray}
These are just the starting relations of the $k$-calculus. Its assumption is that if observer $A$ sends, at time $t_{1}$, a signal to observer $A'$ moving at a constant relative speed $v$, then it is received at $A'$, at time $t'_{1}$, where $t'_{1}=k(v)t_{1}$.  Because of the principle of relativity if $A'$ sends at time $t'_{2}$ a signal to $A$ then it will be received at time $t_{2}$, where $t_{2} = k(v)t'_{2}$.  This is just the result obtained in equation (\ref{eq:tk}).  

Thus it is possible to use the principle of Galilean relativity, the constancy of the speed of light and like-light movements to abstract Minkowski space-time.  Light-like movements are, of course, simply light signals.  For more details of this approach see Kauffman \cite{lk87} \cite{lk01}.

\subsection{The Light Cone Geometry.}

The results of the previous subsection indicates that the basic movements can be factorised into what seem to be a dual pair of spinors, namely, elements of a minimal left ideal.  We now want to show that these elements can be given a meaning in terms of the light cone structures.  This means that the properties of the light cones are implicit in the Clifford algebra itself.

Let us see how this works.  Consider the Pauli Clifford algebra, ${\cal{C}}_{3,0}$ and let us take the idempotent to be $(1 +e_{3})/2$. Then we can form the element
\begin{eqnarray}
\Psi_{L}({\bf r},t) =\psi_{L}({\bf r},t)(1 + e_{3})/2				\label{eq:lr}
\end{eqnarray}
where
\begin{eqnarray*}
\psi_{L}= g_{0}({\bf r},t) + g_{1}({\bf r},t)e_{23} + g_{2}({\bf r},t)e_{31} + g_{3}({\bf r},t)e_{12} 
\end{eqnarray*}
Let us now define a vector ${\cal{V}}$ in the Clifford algebra through the relation ${\cal{V}} =\Psi_{L}\Psi_{R}= \psi_{L}(1+e_{3})\psi_{R}$
where $\psi_{R} = {\widetilde \psi_{L}}$.  Here $\sim$ denotes  the Clifford conjugate.  Then
\begin{eqnarray*}
{\cal{V}} = v_{0}1+ v_{1}e_{1}+v_{2}e_{2}+v_{3}e_{3}
\end{eqnarray*}
where 
\begin{eqnarray}
v_{0}=g_{0}^{2}+g_{1}^{2}+g_{2}^{2}+g_{3}^{2}\hspace{1cm}
v_{1}=2(g_{1}g_{3}-g_{0}g_{2})\hspace{0.5cm}\nonumber \\v_{2}=2(g_{0}g_{1}+g_{2}g_{3})\hspace{1cm}v_{3}=g_{0}^{2}-g_{1}^{2}-g_{2}^{2}+g_{3}^{2}.\hspace{0.3cm}  \label{eq:Hopf}				
\end{eqnarray}
This will be immediately recognised as a Hopf map \cite{mn90}, which is not surprising since we are essentially dealing with light spheres as we will show in section \ref{sec:PFL}.

Furthermore equation (\ref{eq:Hopf}) shows that $v_{0}^{2}= v_{1}^{2}+ v_{2}^{2}+v_{3}^{2}$.  Thus
if we now map $\phi:1 \rightarrow e_{0}$ with $e_{0}^{2}=-1$ and impose the condition $e_{0}e_{i}+e_{i}e_{0} = -2\delta_{0i}$ we have lifted the vector into the larger Clifford algebra ${\cal{C}}_{3,1}$.  However since now $v_{0}^{2}+ v_{1}^{2}+ v_{2}^{2}+v_{3}^{2} = 0$, we have constructed a null vector in this larger Clifford algebra.  If we now project this Clifford vector into a vector in $V_{3,1}$ by $\eta : {\cal{C}}_{3,1} \rightarrow V_{3,1}$ we have constructed a light ray ${\bf v}({\bf r},t)$ in $V_{3,1}$. If we fix the vector at the origin of the co-ordinate system in $V_{3,1}$, then  vary ${\bf r}, t$ we generate a light cone in $V_{3,1}$.  Thus light-like movements in the Clifford algebra can be used to generate a light cone structure on the vector space $V_{3,1}$ which can be taken to be our space-time.  This means that we have used elements of the minimal  left ideal and its conjugate to generate the light cone.

This result is not that surprising because our elements of ${\cal{I}}_{L}$ are simply spinors in another guise. Normally we use matrices to represent  spinors.  Thus in more familiar form
\begin{eqnarray}
|\Psi\rangle\rightarrow
\begin{pmatrix}
     \psi_{1}   \\
      \psi_{2}
\end{pmatrix}
				\label{eq:matspin1}			
\end{eqnarray}
Then we write 
\begin{eqnarray*}
|\Psi\rangle\langle \Psi | = \begin{pmatrix}
     |\psi_{1}|^{2} & \psi_{1}\psi^{*}_{2}   \\
      \psi_{2}\psi^{*}_{1}& |\psi_{2}|^{2}
\end{pmatrix}
\end{eqnarray*}
If we now compare this with the matrix that Penrose introduces to describe a light ray,
\begin{eqnarray}
X=\begin{pmatrix}
     t +z & x-iy   \\
      x + iy& t-z
\end{pmatrix}
					\label{eq:Xtox}
\end{eqnarray}
we find 
\begin{eqnarray*}
t = |\psi_{1}|^{2}+|\psi_{2}^{*}|^{2}\hspace{0.5cm}x = \psi_{1}\psi_{2}^{*}+\psi_{1}^{*}\psi_{2}\hspace{0.5cm}y = i(\psi_{1}\psi_{2}^{*}-\psi_{1}^{*}\psi_{2})\hspace{0.5cm}z = |\psi_{1}|^{2}-|\psi_{2}^{*}|^{2}
\end{eqnarray*}
These relations are, in fact, identical to the expression in \eqref{eq:Hopf} provided first we map $(v_{0},v_{1},v_{2},v_{3})\rightarrow (t, x, y, z)$ and then use the relations between $(g_{0}, g_{1}, g_{2}, g_{3})$ and $(\psi_{1}, \psi_{2})$ presented in Hiley and Callaghan \cite{bhrc08a}.  The mathematical reason for this to work runs as follows. We can map the equivalence class of minimal left ideals onto a Hilbert space $\rho:\Psi_{L}\rightarrow |\psi \rangle$.  In this way we can represent the elements of the left ideal as matrices.  Then $\rho: \Psi_{L}\Psi_{R}\rightarrow |\psi\rangle\langle\psi|$, the expression for the conventional density matrix.

\subsection{The Dirac Clifford and $Sl(2\Bbb{C})$.}

In the previous sub-section although we started from the Pauli Clifford, ${\cal{C}}_{3,0}$, we lifted our structure into the Dirac Clifford ${\cal{C}}_{3,1}$ and found that we have  generated  light cones.  Now it is well known that $Sl(2\Bbb{C})$ is the double cover of $O^{\uparrow}_{+}(1,3)$ \cite{mn90}.  The irreducible representations of $Sl(2\Bbb{C})$ are two-dmensional whereas the Dirac spinor is four-dimensional. The Dirc spinor, while being an irreducible representation of the Clifford algebra, is not an irreducible representation of the spin group $Sl(2\Bbb{C})$. 

 We can express this more formally by considering a Clifford bundle over an $n$-dimensional orientable manifold $M$.  Call the set of sections of this bundle $\Delta(M)$.  Then the Dirac spinor $\Psi_{L} \in \Delta(M)$ is irreducible.  The irreducible representations of the spin group are classified by the eigenvalues of $e_{n+1} = ie_{1}\dots e_{n}$.  In the Dirac case $(e_{n+1})^{2}=1$, so that the eigenvalues of $e_{n+1}$ are $\pm 1$. $\Delta (M)$ is then split into two eigenspaces
 \begin{eqnarray}
\Delta(M)=\Delta^{+}(M)\oplus\Delta^{-}(M)
					\label{eq:2e}
\end{eqnarray}
We can now introduce two projection operators, $P^{\pm}$, defined by
\begin{eqnarray*}
P^{\pm}=(1\pm e_{n+1})
\end{eqnarray*}
so that\begin{eqnarray*}
P^{+}\Psi_{L} = 
\begin{pmatrix}
  \Psi_{L}^{+} \\
  0
\end{pmatrix}\in \Delta^{+}(M),
\hspace{2cm}
P^{-}\Psi_{L} = 
\begin{pmatrix}
  0 \\
  \Psi_{L}^{-}
\end{pmatrix}\in \Delta^{-}(M).
\end{eqnarray*}
In this representation, a general element of the Lorentz group takes the form
\begin{eqnarray*}
L=\begin{pmatrix}
     \Lambda&0   \\
      0&(\Lambda^{\dag})^{-1}
\end{pmatrix} \hspace{2cm}
L^{-1}=\begin{pmatrix}
     \Lambda^{-1}&0   \\
      0&\Lambda^{\dag}
\end{pmatrix}
\end{eqnarray*}
Here $\Lambda$ is the 2 by 2 matrix
\begin{eqnarray*}
\Lambda=\begin{pmatrix}
     \alpha & \beta   \\
      \gamma & \delta
\end{pmatrix}
\end{eqnarray*}
where $\alpha, \beta, \gamma, \delta \in \Bbb{C}$ and $\alpha \delta-\beta \gamma = 1$.  Also
\begin{eqnarray*}
(\Lambda^{\dag})^{-1} = C\Lambda^{*}C^{-1}=\begin{pmatrix}
     \delta^{*} &-\gamma^{*}   \\
     -\beta^{*}   & \alpha^{*}
\end{pmatrix}
\end{eqnarray*}
and
\begin{eqnarray*}
C=\begin{pmatrix}
     0 & 1   \\
      -1 & 0
\end{pmatrix}=-C^{-1}
\end{eqnarray*}
In this way we recognise that $\Lambda$ is the irreducible representation of $Sl(2\Bbb{C})$.  This group has a conjugate irreducible representation, $\Lambda^{*}\in\overline{Sl(2\Bbb{C})}$. These together with the respective dual irreducible representations $(\tilde{\Lambda})^{-1}$ and $(\Lambda^{\dag})^{-1}$ provide all the necessary information to completely describe the role of the Dirac spinor.  Here $\sim$ is the transpose of the matrix representation.

Indeed the spinor defined in equation \eqref{eq:lr} corresponds to $P^{+}\Psi_{L}$ and clearly transforms as 
\begin{eqnarray*}
\Psi^{+\prime }_{L}=\Lambda \Psi^{+}_{L}\hspace{1cm} \mbox{while}\hspace{1cm}\Psi^{-\prime}_{L}= (\Lambda^{\dag})^{-1}
\Psi^{-}_{L}
\end{eqnarray*}
On the other hand we also have $\Psi^{+\prime }_{R}=\Psi^{+}_{R}\Lambda^{\dag}$, where $ \Psi^{+\prime }_{R}=\widetilde{\Psi^{+\prime }_{L}}$.  Here $\sim$ is the Clifford conjugate.

If we now form $\Psi^{+}_{L}\Psi^{+}_{R}=X(x)$ where $X$ is given by equation \eqref{eq:Xtox}.  Then under a Lorentz transformation
$X(x')=\Lambda X(x)\Lambda$, or more transparently 
\begin{eqnarray*}
X(Lx)=\Lambda(L)X\Lambda(L)^{\dag}
\end{eqnarray*}
We can form a dual matrix $\overline{X}(x)$ which is constructed from $\psi_{L}(1-e_{3})\psi_{R}$.  This means that in the matrix representation used above
\begin{eqnarray*}
\overline{X}=\begin{pmatrix}
     t -z & -x+iy   \\
      -x - iy& t+z
\end{pmatrix}
\end{eqnarray*}
This gives, under a Lorentz transformation
\begin{eqnarray*}
\overline{X}(Lx)=(\Lambda^{\dag})^{-1}\overline{X}(x)\Lambda^{-1}
\end{eqnarray*}
It may be of some interest to note the following results:
\begin{eqnarray*}
\overline{X}(x)X(y)+\overline{X}(y)X(x)=2{\bf x.y}\\
\overline{X}(x)X(y)-\overline{X}(y)X(x)=2X(x^{0}{\bf y}-y^{0}{\bf x}+{\bf x\wedge y})\\
detX(x)=\overline{X}(x()X(x)={\bf x}^{2}\\
\overline{X}X(y)-X(y)\overline{X}(x)=2X({\bf x\wedge y}).
\end{eqnarray*}

Finally let us return to the matrix representation of the Dirac algebra so that we can fit our approach into the more conventional approach.  We write
\begin{eqnarray*}
\Psi=\begin{pmatrix}
     \lambda   \\
      \rho
\end{pmatrix}
\end{eqnarray*}
where $\lambda$ and $\rho$ are column spinors of dimension 2.  In fact
\begin{eqnarray*}
\Psi^{+}_{L}\rightarrow\lambda \hspace{2cm}\Psi^{-}_{L}\rightarrow \rho.
\end{eqnarray*}
The two spinors $\lambda$ and $\rho$ correspond to the chiral Weyl spinors which are used to describe the left- and right-handed neutrino states.  But notice that in our approach we have nowhere introduced quantum mechanics. The Clifford algebra is simply a way to describe the geometry of space-time.  What is remarkable is the way physicists are introduced to the Dirac formalism. It appears that we are forced into the Clifford algebra by quantum mechanics, but this is manifestly not the case.  Rather we should introduce the algebra from classical geometry and then show that quantum mechanics exploits the structure of this geometry.  In this way quantum mechanics is not all that strange or novel.

Formally these ideas can be put into the language of fibre bundles.  The left-handed Weyl spinor is a section of the spin bundle $(W, \pi, M, \Bbb{C}^{2}, Sl(2\Bbb{C}))$ while the right-handed Weyl spinor is a section of $(\overline{W}. \pi, M, \Bbb{C}, \overline{Sl(2\Bbb{C})})$.  In this language the Dirac spinor is a section of $(D, \pi, M, \Bbb{C}, Sl(2\Bbb{C})\oplus\overline{Sl(2\Bbb{C})})$.

\subsection{Past and Future Light cones.}
\label{sec:PFL}

Consider a null ray defined by
\begin{eqnarray*}
{\bf r}^{2}-c^{2}t^{2}=0
\end{eqnarray*}
Then for a fixed $t$, the light cone intersects a space-like hypersurface in a sphere with raduis $r=\pm ct$.  This means that if $t$ is positive, the sphere is on the future light cone, while if $t$ is negative, the sphere is on the past light cone.  However the points on a sphere can be characterised by a stereographic projection onto points on a plane with coordinates $\lambda=(\xi, \eta)$ \cite{ffbh81}.  This we write as a matrix 
\begin{eqnarray}
\psi=\begin{pmatrix}
     \xi   \\
      \eta
\end{pmatrix}
				\label{eq:WeylSpinor}
\end{eqnarray}
Under a Lorentz transformation we have
\begin{eqnarray}
\psi' =\Lambda_{1}(L)\psi
					\label{eq:spintrans}
\end{eqnarray}
Now we have seen that
\begin{eqnarray*}
x_{0}=|\xi|^{2}+|\eta|^{2}, \qquad x_{1}=\xi\bar{\eta} +\bar{\xi}\eta \qquad x_{2}=i(\xi\bar{\eta} -\bar{\xi}\eta)\qquad x_{3}=
 |\xi|^{2}-|\eta|^{2}
\end{eqnarray*}
In the dotted and undotted notation of Penrose and Rindler \cite{rpwr84} these results follow from
\begin{eqnarray*}
x_{\mu}=\sigma_{\mu}^{A\dot{A}}\psi_{A}\bar{\psi}_{\dot{A}}
\end{eqnarray*}
where $\sigma_{\mu}^{A\dot{A}}$ are the spin frames
\begin{eqnarray*}
\sqrt{2}\sigma_{0}^{A\dot{A}}=\begin{pmatrix}
     1 & 0   \\
      0 & 1
\end{pmatrix}\quad\sqrt{2}\sigma_{1}^{A\dot{A}}=\begin{pmatrix}
     0 & 1   \\
      1 & 0
\end{pmatrix}\quad\sqrt{2}\sigma_{2}^{A\dot{A}}=\begin{pmatrix}
     0 & i  \\
      -i & 0
\end{pmatrix}\quad\sqrt{2}\sigma_{3}^{A\dot{A}}=\begin{pmatrix}
     1 & 0   \\
      0 & -1
\end{pmatrix}\quad
\end{eqnarray*}
Here we identify equation (\ref{eq:WeylSpinor}) with $\psi_{A}$ via $\psi_{1}=\xi$ and $\psi_{2} = \eta$.

Let us look at the infinitesimal version of the transformation (\ref{eq:spintrans}) by  writing
\begin{eqnarray*}
\Lambda_{1}(L) = 1 +\epsilon\begin{pmatrix}
     \alpha & \beta   \\
      \gamma & \delta
\end{pmatrix}\quad
\end{eqnarray*}
then
\begin{eqnarray*}
\Delta\xi=\epsilon(\alpha\xi+\beta\eta)\qquad
\Delta\eta=\epsilon(\gamma\xi+\delta\eta)\\
\Delta\bar{\xi}=\epsilon(\bar{\alpha}\bar{\xi}+\bar{\beta}\bar{\eta})\qquad
\Delta\bar{\eta}=\epsilon(\bar{\gamma}\bar{\xi}+\bar{\delta}\bar{\eta})
\end{eqnarray*}
Thus $x_{\mu}$ transforms as
\begin{eqnarray*}
x'_{\mu} = 1 + \omega^{\nu}_{\mu}x_{\nu}
\end{eqnarray*}
with
\begin{eqnarray}
\omega^{\nu}_{\mu}=\begin{pmatrix}
     0 & -a_{4}&-a_{5}&-a_{6}   \\
      -a_{4}&0&a_{3}&-a_{2}\\
      -a_{5}&-a_{3}& 0& a_{1}\\
      -a_{6}&a_{2}&-a_{1}&0
      \end{pmatrix}
      					\label{eq:ilt}
\end{eqnarray}
where
\begin{eqnarray}
2a_{1}=i(-\beta+\bar{\beta}-\gamma+\bar{\gamma})\qquad
2a_{2}=(\beta+\bar{\beta}-\gamma-\bar{\gamma})\qquad
2a_{3}=i(-\alpha+\bar{\alpha}+\delta-\bar{\delta})\nonumber\\
2a_{4}=(-\beta-\bar{\beta}-\gamma-\bar{\gamma})\qquad
2a_{5}=i(-\beta+\bar{\beta}+\gamma-\bar{\gamma})\qquad
2a_{6}=(-\alpha-\bar{\alpha}+\delta+\bar{\delta}).
					\label{eq:slilt}
\end{eqnarray}

Now let us examine the past light cone.  Let us again use the stereographic projection only this time using coordinates $\rho =(\sigma, \tau)$ which we can again write this in matrix form
\begin{eqnarray}
\phi=\begin{pmatrix}
     \sigma   \\
      \tau
\end{pmatrix}.
					\label{eq:WeylSpinor2}
\end{eqnarray}
Under a Lorentz transformation we have
\begin{eqnarray*}
\phi' = \Lambda_{2}(L)\phi.
\end{eqnarray*}
Then since the infinitesimal Lorentz transformation can be written as
\begin{eqnarray*}
\Lambda_{2}(L) = 1+\epsilon\begin{pmatrix}
     A & B   \\
      \Gamma & \Delta\end{pmatrix},
\end{eqnarray*}
we find
\begin{eqnarray*}
\Delta\sigma=\epsilon(A\sigma+B\tau)\qquad
\Delta\tau=\epsilon(\Gamma\sigma+\Delta\tau)\\
\Delta\bar{\sigma}=\epsilon(\bar{A}\bar{\sigma}+\bar{B}\bar{\tau})\qquad
\Delta\bar{\tau}=\epsilon(\bar{\Gamma}\bar{\sigma}+\bar{\Delta}\bar{\tau}).
\end{eqnarray*}

Let us see what happens to a vector that can be written in the form
\begin{eqnarray*}
y_{0}=-(|\sigma|^{2}+|\tau|^{2}), \qquad y_{1}=\sigma\bar{\tau} +\bar{\sigma}\tau \qquad y_{2}=i(\sigma\bar{\tau} -\bar{\sigma}\tau)\qquad y_{3}=
 |\sigma|^{2}-|\tau|^{2}
\end{eqnarray*}
This becomes in the dotted-undotted notation
\begin{eqnarray*}
y^{\mu}=\sigma_{\dot{A}A}^{\mu}\psi^{A}\overline{\psi}^{\dot{A}}
\end{eqnarray*}
where we can identify $\overline{\psi}^{\dot{A}}$ with $\phi$ in equation \eqref{eq:WeylSpinor2} through the relations $\overline{\psi}^{\dot{1}}=\sigma$ and $\overline{\psi}^{\dot{2}}= \tau$.  Writing
\begin{eqnarray*}
y^{\prime\mu}=1+\omega^{\mu}_{\nu}y^{\nu}
\end{eqnarray*}
and comparing this with the expression \eqref{eq:ilt}, we find
\begin{eqnarray*}
2a_{1}=i(-B+\bar{B}-\Gamma +\bar{\Gamma})\qquad
2a_{2}=(B+\bar{B}-\Gamma\bar{\Gamma})\qquad
2a_{3}=i(-A+\bar{A}+\Delta-\bar{\Delta})\\
2a_{4}=(B+\bar{B}+\Gamma+\bar{\Gamma})\qquad
2a_{5}=i(B+\bar{B}-\Gamma+\bar{\Gamma})\qquad
2a_{6}=(A+\bar{A}-\Delta-\bar{\Delta})
\end{eqnarray*}
Comparing this expression with the expressions given in \eqref{eq:slilt} we find
\begin{eqnarray*}
A=\bar{\delta}\qquad B=- \bar{\gamma}\qquad \Gamma =- \bar{\beta}\qquad \Delta =\bar{\alpha}
\end{eqnarray*}
This means that the expression for $\Lambda_{2}$ becomes
\begin{eqnarray*}
\Lambda_{2}=1 + \epsilon\begin{pmatrix}
     \bar{\delta} & -\bar{\gamma}   \\
      -\bar{\beta} & \bar{\alpha} 
\end{pmatrix}
\end{eqnarray*}
Thus 
\begin{eqnarray*}
\Lambda_{2}(L)=(\Lambda^{*})^{-1}
\end{eqnarray*}
This means that $\Lambda_{1}\in Sl(2\Bbb{C})$ and $\Lambda_{2}\in \overline{Sl(2\Bbb{C})}$.  With these results we can then argue that the spinor $\psi_{A}$ can be used to describe the future light cone, while $\overline{\psi}^{\dot{A}}$ describes the past light cone.

Since in the matrix representation the Dirac spinor can be written as
\begin{eqnarray*}
\Psi=\begin{pmatrix}
      \psi_{A}   \\
      \overline{\psi}^{\dot{A}}  
\end{pmatrix}
\end{eqnarray*}
 we see the Dirac spinor contains information about the future and past light cones. 
 
 If we return to the Clifford algebra itself rather than a matrix representation, an element of the minimal left ideal of the Dirac Clifford $\Psi_{L}$ contains the information necessary to describe both the future and past light cones.
 
 It is interesting to note that we are able to show how $\psi_{A}$ and $\overline{\psi}^{\dot{A}}$ are related to stereographic projections of a point on the light sphere emerging from the origin $O$.  We illustrate this relationship in figure 4.  The light ray from $O$ passes through the light cone at the point $P$.  If we project this point from the North pole, $N$, to the point $Q_{N}$ we find the spinor co-ordinates of this point is $\overline{\psi}^{\dot{A}}$.  If we project $P$ from the South pole, $S$, to the point $Q_{S}^{d}$ we find the spinor co-ordinates of this point is $\psi_{A}$.  Under a Lorentz transformation $Q_{N}$ and $Q_{S}^{d}$ transform together.  For more details on stereographic projections see Frescura and Hiley \cite{ffbh81}.

 
 \begin{figure}[h]
 \begin{center}
 \includegraphics[width=5in]{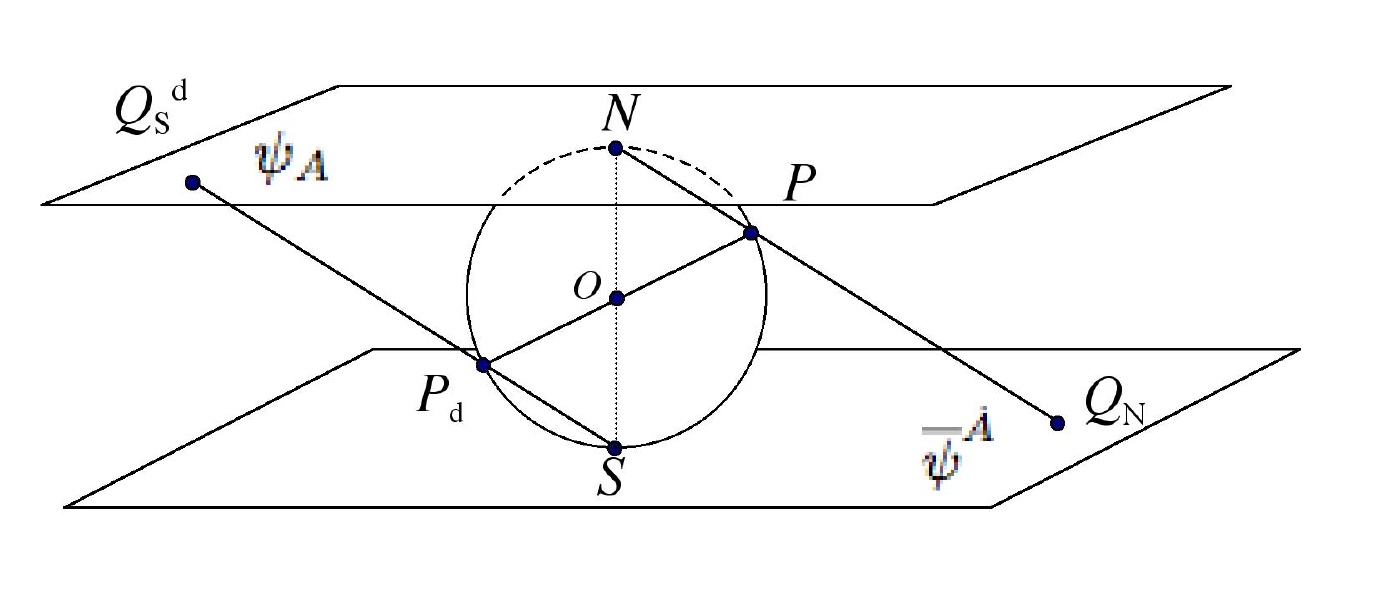}
 \caption{Spinors describing stereographic projections  }
 \end{center}
 \end{figure}

We can discuss all of this in terms of elements of the minimal ideals, which begins to explain why  the Hopf transformation in equation \eqref{eq:Hopf} appears.  To do this we first note that $\Psi_{L}$ contains elements from a pair of minimal left ideals of the covering group, one  corresponds to $Sl(2\Bbb{C})$ itself, the other corresponds to $\overline{Sl(2\Bbb{C})}$ as we have discussed above.  In fact if we introduce the projection operators $(1 \pm ie_{5})/2$ we find that $\psi_{A}$ has components
 \begin{eqnarray*}
(1-ie_{5})(1-e_{03})/4\qquad\mbox{and}\qquad(1-ie_{5})(e_{13}+e_{01})/4.
\end{eqnarray*}
The other element that we are interested in is $\bar{\psi}^{\dot{A}}$ which has components
\begin{eqnarray*}
(1+ie_{5})(e_{0}-e_{3})/4\qquad\mbox{and}\qquad(1+ie_{5})(e_{2}+e_{023})/4.
\end{eqnarray*}
These projections give rise to the left- and right-handed helicity or Weyl states \cite{lr85}.  The notation can become more physically meaningful if we write
\begin{eqnarray*}
\Psi=\begin{pmatrix}
      \psi_{\lambda}   \\
     \psi_{\rho}  
\end{pmatrix},
\end{eqnarray*}
where the suffixes $\lambda$ and $\rho$ denote the left- and right-handed Weyl spinors respectively.

It should be noted that so far all this discussion has been about {\em classical} space-time.  We have done absolutely nothing to suggest that this mathematics has anything to do with quantum theory. Let us continue in this vein by going to the conformal Clifford ${\cal{C}}_{2,4}$ where we will find the twistor of Penrose \cite{rp67} appearing.

\section{The conformal Clifford ${\cal{C}}_{2,4}$}

\subsection{The Twistor and Light-cone Structures.}
\label{sec:TLS}

In order to explain how the twistor plays a role in relating different light cones at different positions in space-time, we will not start from the structure of elements of the minimum left ideals but  from a specific matrix representation.  This will help us to see how this bigger structure is related to the Dirac spinors discussed in section \ref{sec:PFL}

We begin by considering a six-dimensional hypersphere given by the equation
\begin{eqnarray*}
(\beta_{A} \xi^{A})^{2} = \Omega^{2}\hspace{1cm}\mbox{where}\hspace{1cm} A=(0, 1, 2, 3, 4, 5)
\end{eqnarray*}
Here we have departed from our standard notation by using $\beta$s instead of $e$s for the generators of the Clifford algebra.  Thus we write $\{\beta_{A}\}$ for the generators of the conformal Clifford algebra ${\cal{C}}_{2,4}$. We choose to write the metric tensor in the form $g_{AB}=(+----+)$ so that $[\beta_{A}, \beta_{B}]=g_{AB}$.  The spinors of the six-space, $\xi^{A}$, are then defined through the relationship
\begin{eqnarray}
(\beta_{A}\xi^{A})\Psi=0
					\label{eq:bitwist}
\end{eqnarray}
Now we will choose a specific representation of the generators such that 
\begin{eqnarray}
\beta_{\mu}= 1\otimes\gamma_{\mu}\hspace{1cm}\beta_{4}=\sigma_{1}\otimes\gamma_{5}\hspace{1cm}\beta_{5}=\sigma_{1}\sigma_{3}\otimes\gamma_{5}\nonumber
\end{eqnarray}
\begin{eqnarray}
\beta_{i}^{2}=-1;\hspace{1cm}\beta_{0}^{2}=1;\hspace{1cm}
\beta_{4}^{2}=-1;\hspace{1cm}\beta_{5}^{2}= 1.\nonumber
\end{eqnarray}
Here the $\gamma_{\mu}$ are the generators of the Dirac Clifford with $\mu = \{0, 1, 2, 3\}$.  The $\sigma$s are the generators of the Pauli Clifford algebra.  By writing things in this way, it becomes clear how these algebras are related to each other.
 Equation \eqref{eq:bitwist} becomes
\begin{eqnarray}
\begin{array}{c}(\xi^{4}+\xi^{5})\psi_{\lambda_{2}}=-i(\xi^{0}-\sigma_{i}\xi^{i})\psi_{\rho_{1}}
\\(\xi^{4}-\xi^{5})\psi_{\lambda_{2}}=-i(\xi^{0}-\sigma_{i}\xi^{i})\psi_{\rho_{2}}\\
(\xi^{4}+\xi^{5})\psi_{\rho_{2}}=i(\xi^{0}+\sigma_{i}\xi^{i})\psi_{\lambda_{1}}
\\(\xi^{4}-\xi^{5})\psi_{\rho_{1}}=i(\xi^{0}+\sigma_{i}\xi^{i})\psi_{\lambda_{2}}
\end{array}\hspace{0.5cm}
\mbox{where}\hspace{0.5cm}\Psi=\left(\begin{array}{c} \psi_{\lambda_{1}}\\\psi_{\lambda_{2}}\\\psi_{\rho_{1}}\\\psi_{\rho_{2}} \end{array}\right)
					\label{eq:xiInpsi}
\end{eqnarray}
where $\Psi$ is a bi-twistor which has eight components.  The subscripts  $\lambda$ and $\sigma$ again signify the left- and right-handed Weyl spinors, only now we have two pairs of Weyl spinors.   Thus the bi-twistor contains of two Dirac four-spinors
\begin{eqnarray}
\left(\begin{array}{c}\psi_{\lambda_{1}}\\0\\\psi_{\rho_{1}}\\0\end{array}\right)
\hspace{1cm}\mbox{and}\hspace{1cm}\left(\begin{array}{c}0\\\psi_{\lambda_{2}}\\0\\\psi_{\rho_{2}}\end{array}\right)
						\label{eq:2lightcones}
\end{eqnarray}			 
This means that we can describe two light cones within a single bi-twistor.  We will see that it is this feature of the Penrose bi-twistor that  will enable us to relate light cones at different points.

We have shown elsewhere \cite{bh08} that the bi-twistor also contains a pair of Penrose twistors which can be written in the form
\begin{eqnarray}
\left(\begin{array}{c}\psi_{\lambda_{1}}\\0\\0\\\psi_{\rho_{2}}\end{array}\right)
\hspace{1cm}\mbox{and}\hspace{1cm}\left(\begin{array}{c}0\\\psi_{\lambda_{2}}\\\psi_{\rho_{1}}\\0\end{array}\right)\nonumber
\end{eqnarray}
As we will see the doubling will give a symmetrical relation between the pair of light cones described in equation (\ref{eq:2lightcones}).

Thus we see that there is a deep relationship between the Dirac spinors and the Penrose twistors which is sometimes overlooked in the component notation normally used to discuss twistors \cite{rp67}.  Furthermore since the Clifford algebra is a geometric algebra, this relationship must be geometric in nature and we emphasise, once again, that this has nothing {\em per se} to do with quantum mechanics.  In other words the conformal spinor can be regarded as a combination of various sub-spinors and we need to find out the exact meaning of these spinors. 

 To do this let us go on to consider the relation between the conformal structure and Minkowski space.  We will see that this is a projective relationship, which entails breaking the symmetry by fixing the light cone at infinity.
To bring this feature out, we need to know how the co-ordinates $x^{\mu}$ of a point in Minkowski space-time are related to the projective co-ordinates $\xi^{A}$ in the six-dimensional hyperspace.  The projective relationship we need is 
 \begin{eqnarray}
x^{i}=\frac{\xi^{i}}{\xi^{4}+\xi^{5}};
\hspace{1cm}\mbox{and}\hspace{1cm}
x^{0}=\frac{\xi^{0}}{\xi^{4}+\xi^{5}}.
\end{eqnarray}
This means that at the origin of our co-ordinate system in Minkowski space, we have $x^{0}=x^{i}=\xi^{0}=\xi^{i}=0$.   We need one more relationship  to completely specify the origin in terms of the $\xi^{A}$.  This relationship is
\begin{eqnarray}
\left(\xi^{4}\right)^{2}-\left(\xi^{5}\right)^{2}=0=\left(\xi^{4}-\xi^{5}\right)\left(\xi^{4}+\xi^{5}\right)
\end{eqnarray}
Thus if $\left(\xi^{4}+\xi^{5}\right)\neq0$ then $\left(\xi^{4}-\xi^{5}\right)=0$. In terms of the projective co-ordinates, the origin is specified by
\begin{eqnarray}
x^{0}=x^{i}=\xi^{0}=\xi^{i}=\left(\xi^{4}-\xi^{5}\right)=0\nonumber
\end{eqnarray}

To see the implications for the spinors discussed above, lets us return to equation (\ref{eq:xiInpsi}) where the spinors $(\psi_{\lambda},\psi_{\rho})$   are expressed in terms of $\xi^{A}$.  If we examine the first and third equation in (\ref{eq:xiInpsi}) we see immediately that $\psi_{\lambda_{2}}=\psi_{\rho_{2}}=0$.  This means that the bi-twistor at the origin is identified by
\begin{eqnarray}
\Psi=\left(\begin{array}{c}\psi_{\lambda_{1}}\\0\\\psi_{\rho_{1}}\\0\end{array}\right)\nonumber
\end{eqnarray}
Thus this bi-twistor represents light rays on the light cone at the origin.  

To see the meaning of the second  term in equation (\ref{eq:2lightcones}) we must first recall the expression for the translation operator in the conformal group.  This can be shown to be \cite{bh08}
\begin{eqnarray}
U(\Delta x)=1-\Delta x^{\mu}P_{\mu}
\hspace{0.5cm}\mbox{where}\hspace{0.5cm}
P_{\mu}=\frac{1}{2}\beta_{\mu}(\beta_{4}-\beta_{5})\nonumber
\end{eqnarray}
Here $\Delta x$ is the displacement produced.

\begin{figure}[h]
\begin{center}
\includegraphics[width=3in]{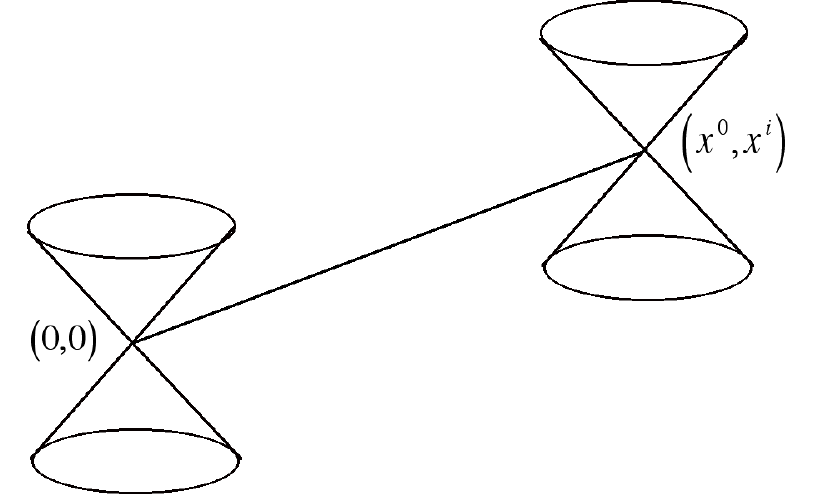}
\end{center}
\caption{ Relative location of light cones.}
\end{figure}

 Applying this to the bi-twistor we find
\begin{eqnarray}
U(\Delta x)\left(\begin{array}{c}\psi_{\lambda_{1}}\\\psi_{\lambda_{2}}\\\psi_{\rho_{1}}\\\psi_{\rho_{2}}\end{array}\right)
\hspace{0.5cm}=
\hspace{0.5cm}\left(\begin{array}{c}\psi_{\lambda_{1}}\\-i(\Delta x^{0}-\sigma\Delta {\bf x})\psi_{\rho_{1}}\\\psi_{\rho_{1}}\\i(\Delta x^{0}+\sigma\Delta {\bf x})\psi_{\lambda_{1}}\end{array}\right)
\end{eqnarray}
which means that we have generated a new light ray on a new light cone at $(x^{0},x^{i})$   given by
\begin{eqnarray}
\psi_{\lambda_{2}}=-i(x^{0}-\sigma_{i}x^{i})\psi_{\rho_{1}}
\hspace{0.5cm} \mbox{and}\hspace{0.5cm}
\psi_{\rho_{2}}=-i(x^{0}+\sigma_{i}x^{i})\psi_{\lambda_{1}}\nonumber
\end{eqnarray}
Thus our bi-twistor describes light rays on two light cones, one at the origin of the co-ordinates while the other is at some other point $(x^{0},x^{i})$   as shown in figure 6.

\begin{figure}[h]
\begin{center}
\includegraphics[width=4in]{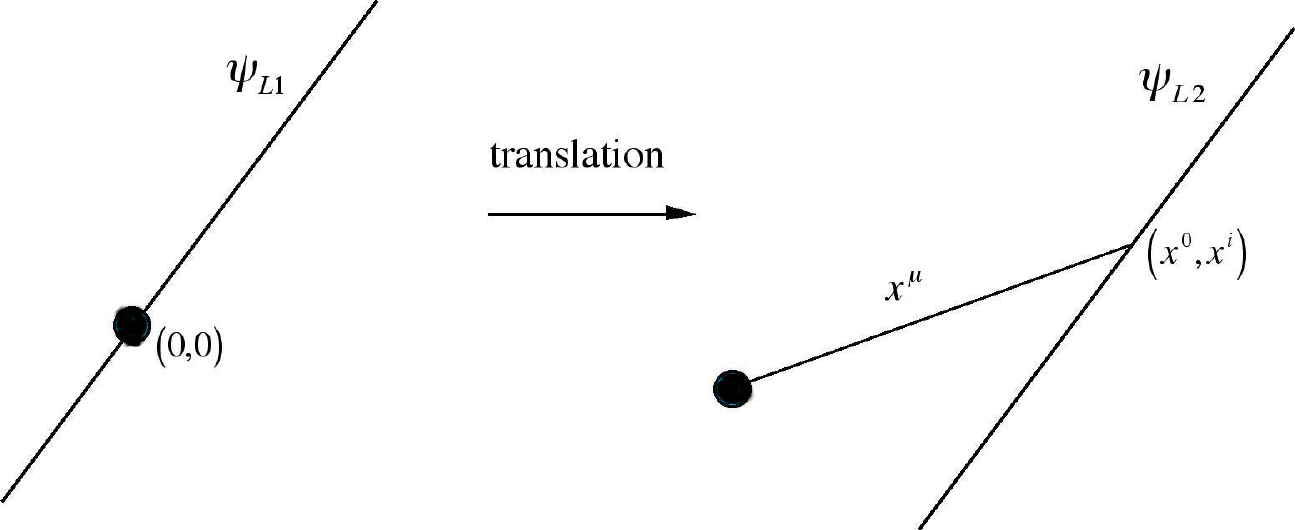}
\end{center}
\caption{ Twistor translates light ray through the origin.}
\end{figure}

Recalling that the origin has $\psi_{\lambda_{2}} = \psi_{\rho_{2}} = 0$, we get the pairing
\begin{eqnarray}
\left(\begin{array}{c}\psi_{\lambda_{1}}\\ i(\Delta x^{0}+\sigma. \Delta {\bf x})\psi_{\lambda_{1}}\end{array}\right)
\hspace{0.5cm}\mbox{and}\hspace{0.5cm}
\left(\begin{array}{c}\psi_{\rho_{1}}\\ -i(\Delta x^{0}-\sigma.\Delta{\bf x})\psi_{\rho_{1}}\end{array}\right) \nonumber
\end{eqnarray}
These are the two twistors that Penrose has introduced, one based on the forward light cone, the other on the past light cone.

All of these results can be obtained from the conformal Clifford using elements of the left ideals.  However we will not discuss these here but will be presented elsewhere.

\subsection{Conclusions Drawn from the Orthogonal Clifford Algebras.}

Let us now summarise the position we have reached so far.  We have shown that the orthogonal Clifford algebras carry a rich light-cone structure and that this structure can be abstracted from the algebra itself without the need to start from an {\em a priori} given vector space.  This means that we do not have to start with a tensor structure on a vector space only to find later that we are forced to introduce the spinor structure, almost as an after thought, a feature that puzzled Eddington. He summed this up by famously remarking, `Something has slipped through the net' \cite{ck94}. Rather we can start by taking process as basic and from the process itself, abstract a light-cone structure.  Key to this approach is the interpretation of the structure of elements of the minimal left and right ideals of the algebra.

Recall that we started by considering an undivided whole in the form of the holomovement.  We then followed Kauffman's \cite{lk98} by making  a mark or distinction in this total flux of process and then define a series of `extensons', which could be ordered into a groupoid.  By adding structure, this groupoid can be developed into an algebra.  We concentrated on the  orthogonal symmetry that we see around us and showed how to construct  a hierarchy of orthogonal Clifford  algebras.  We then show that from this algebra, we can abstract some of the usual properties of space-time by identify elements of the minimal left ideals with a light-cone structure. In fact we arrive at a structure of light-cones which ensures the underlying manifold we have constructed  is Poincar\'{e} invariant.  

In doing this we have identified the elements of a minimal left ideal with the spinors used in the usual approach.  However we have a richer structure since we have sets of  primitive idempotents  from which we generate additional spinor structures, but in the usual approach these are all treated as equivalent and therefore loose the possibility of exploiting this additional structure.  One feature of this richer structure is that we are able to identify different distinctions with different idempotents, thus enabling us to obtain an insight to the implicate order introduced by Bohm \cite{db80} as we will explain later.  There is a further advantage of our approach and this removes Eddington's worry because our resulting structure contains {\em both} spinors and tensors on an equal footing since both entities arise naturally within the algebra itself.

Let us look more carefully at how the distinctions are made and how each distinction `carves out' a specific space-time structure.  In the Dirac algebra we make a distinction by picking out a specific time frame which is done by choosing a primitive idempotent $\epsilon =(1+ e_{0})/2$.  Using this idempotent we can then construct the light-cone structure based on this choice of idempotent.  

Another observer can choose another distinction to produce a different time frame, namely, $\epsilon '=(1+e'_{0})/2$.  This will then produce a different light-cone structure.  Of course these structures  are  equivalent since they are related through 
\begin{eqnarray*}
\epsilon ' = g\epsilon g^{-1}, \hspace{1cm}\mbox{where}\hspace{1cm}g \in G.
\end{eqnarray*}
Here $G$ is the Clifford group, which, in the case of the Dirac algebra, is simply the covering group of the Lorentz group.

The reaction to these ideas from the physicist brought up in the traditional way will probably be ``Surely this is just the same as choosing a specific Lorentz  frame of reference."  Of course {\em mathematically} the two approaches are equivalent, but they are based on very different intuitive ideas.  In the approach we are proposing here, the activity is going on in a structure in which there is no distinction between space and time.  In fact, strictly there is no space or time.  There is only a {\em pre-space-time}.  Space-times are {\em constructed} by breaking the symmetry in the total process by choosing a specific idempotent is used to describe the process.   In this special case a specific idempotent to define a specific time-frame.  In other words the idempotents define equivalence classes of observers, namely Lorentz observers.

Perhaps figure 7 may help to bring the idea out more clearly.  Speaking loosely we could say the activity is `going on' in the `space' $E$, whereas our description of what is going on is in the projections $\eta_{i}$, in what I have elsewhere \cite{bh01a} called `shadow manifolds'.  

\begin{figure}[h]
\begin{center}
\includegraphics[width=4in]{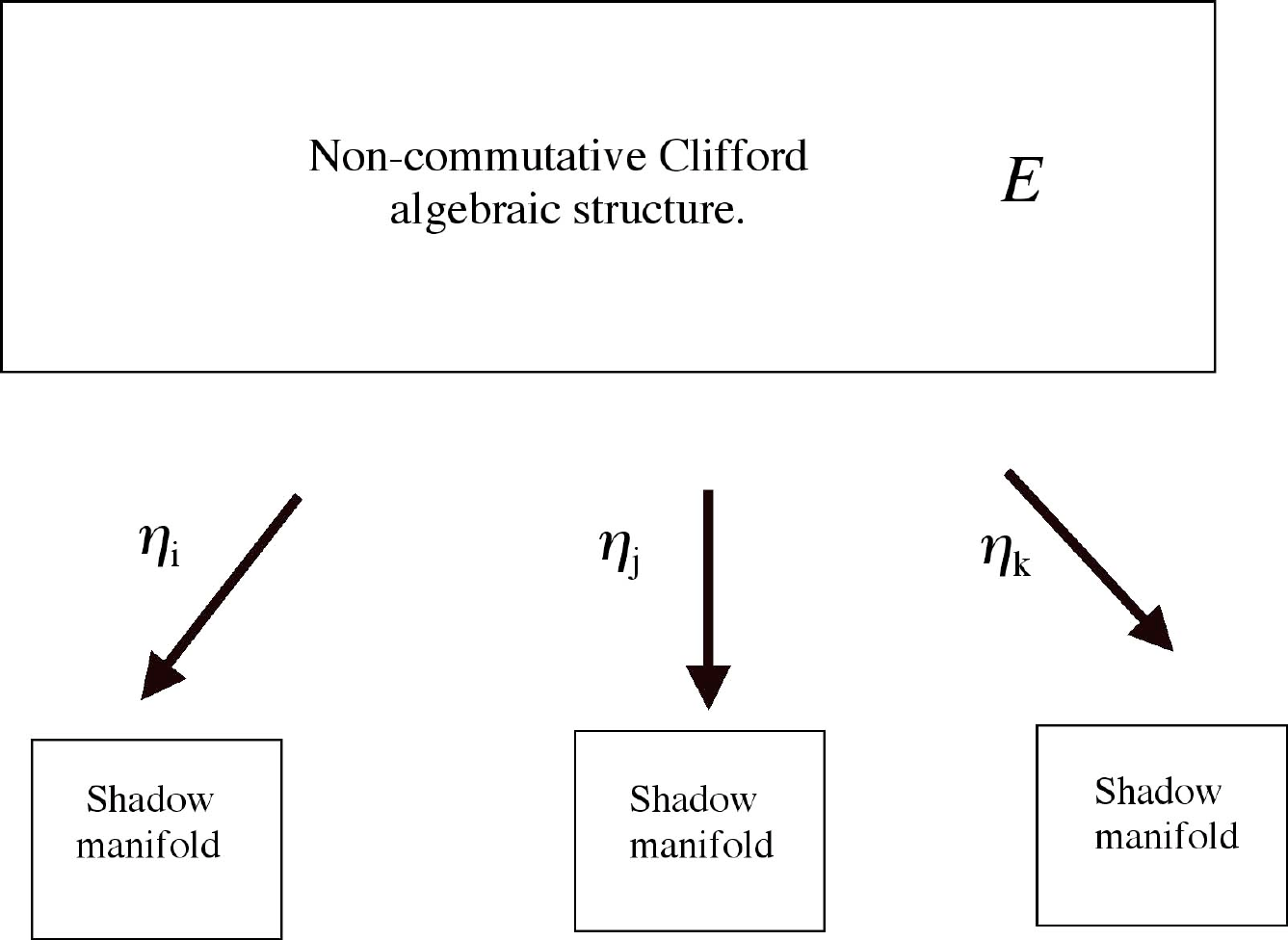}
\caption{The Construction of Shadow Manifolds.  }
\end{center}
\end{figure}

In the language that David Bohm used \cite{db80}, we can regard $E$ as the holomovement or implicate order.  The shadow manifolds are then the sets of explicate orders consistent with the implicate order.  Once again we stress  that the structure we have been discussing is purely classical and contains no feature of quantum mechanics even though we have a non-commutative algebra.  To introduce quantum processes we must add more structure.

\section{Connections in the Clifford Bundle.}

\subsection{The Dirac Connection.}

In the previous section we have ended up with a mathematical structure which has remarkable similarities to a Clifford bundle. If we take just one of the shadow manifolds to be the base space then in fact we have a Clifford bundle.  In this section we will develop our ideas further by exploiting known features of a Clifford bundle \cite{hbl89}.  

We will begin by considering a single shadow manifold and assume it to be a Riemannian manifold, $M$.  We then construct a Clifford bundle ${\cal{C}}_{n,m}(M)$ with this manifold as the base space.   Although not necessary, we will assume the manifold to be flat.  Let $E$ be a bundle of left modules over ${\cal{C}}_{n,m}(M)$ so that at each point $x \in M$, the fibre is a left module over ${\cal{C}}_{n,m}(M)_{x}$. 

 In Section \ref{sec:TLS} we introduced a notion of space  translation which led to the concept of the twistor.  This looked after the geometric properties of the light cone structure.  Now we want to consider particle dynamics so we introduce a first order differential operator $D: \Gamma(E)\rightarrow\Gamma(E)$, where $\Gamma(E)$ is the space of smooth cross-sections of $E$.  The derivative $D$ defines a connection on $M$. If $\sigma\in \Gamma(E)$ then we define the connection through
\begin{eqnarray*}
D\sigma = \sum _{j=1}^{n}e_{j}\partial_{x_{j}}
\end{eqnarray*}
where the $\{e_{j}\}$ are the generators of ${\cal{C}}_{n,m}$ satisfying the relation $\eta(e_{j})=x_{j}$, $x_{j}$ being the local co-ordinates of the point $x$ on $M$.  This connection is called the Dirac operator because it is easy to show that $D^{2}$ is the Dirac Laplacian $\nabla^{2}$.  Notice this definition can be used for {\em any} Clifford algebra and is not restricted to the relativistic algebra ${\cal{C}}_{1,3}$.

It is through this connection that we are able to handle the hierarchy of Clifford algebras, ${\cal{C}}_{0,1}, {\cal{C}}_{3,0}, {\cal{C}}_{1,3}$.  We will call these algebras  the Schr\"{o}dinger, Pauli and Dirac Clifford algebras respectively.  The reason for introducing the term `Schr\"{o}dinger' algebra is because  ordinary non-relativistic quantum mechanics forms a trivial complex line bundle $E=\Bbb{R}^{3}\otimes\Bbb{C}$.  Here the wave function $\psi(x)$ is simply a section of $E$.  This translates into our structure because ${\cal{C}}_{0,1}\cong\Bbb{C}$.  

Notice as we remarked above, each algebra in the hierarchy has its own Dirac operator.  These are 
\begin{eqnarray*}
D&=e\sum_{i=1}^{3}\partial_{x_{i}}\hspace{1cm}\mbox{Schr\"{o}dinger}\\
D&=\sum_{i=1}^{3}\sigma_{i}\partial_{x_{i}}\hspace{1cm}\mbox{Pauli}
\hspace{1cm}\\
D&=\sum_{\mu =1}^{3}\gamma_{\mu}\partial x^{\mu}\hspace{1cm}\mbox{Dirac}\hspace{1cm}
\end{eqnarray*}
We should immediately recognise these as the momentum operators used in quantum mechanics.  Notice the sum does not include the time derivative.  The reason for this will become clear in the next sub-section.

Now it is also possible to view ${\cal{C}}_{n,m}(M)$ as a bundle of right modules which means we can exploit right multiplication, i.e. multiplication from the right.  This means we also have the possibility of a `right' Dirac operator which we will denote as ${\overleftarrow D}=\sum_{j=1}^{n}{\overleftarrow\partial}_{x_{j}}e_{j}$.  The `left' Dirac operator introduced above will be denoted by $\overrightarrow D$. \footnote{It should also be noted that these derivatives are related to the sum and differences of the  boundary, $\delta^{*}$, and co-boundary, $\delta$ operators in the exterior calculus.}
  Notice that 
${\overleftarrow D}^{2}$ also produces the Dirac Laplacian.

\subsection{Relevance to Quantum Mechanics.}

In the conventional approach it is the wave function that is singled out for special attention because it is assumed to contain all the information for describing the state of the system.  We do not have a wave function.  Instead we use elements of the minimal left and right ideals which we single out for special attention.  However as we have pointed out earlier,  the information contained in an element of the left ideal, say $\Psi_{L}$, is exactly the same as that contained in the conventional wave function.

We have already pointed out that the key element in our description is not  $\Psi_{L}$ alone but $\hat{\rho}=\Psi_{L}\Psi_{R}$.  As we have pointed out above $\hat{\rho}$ is equivalent to the density operator for a pure state in the conventional quantum theory.  This means we must consider derivatives of the form $({\overrightarrow D}\Psi_{L})\Psi_{R}$ and $\Psi_{L}({\overleftarrow D}\Psi_{R})$.
Rather than treat these two derivatives separately we will consider expressions like 
\begin{eqnarray*}
({\overrightarrow D}\Psi_{L})\Psi_{R}+\Psi_{L}(\Psi_{R}{\overleftarrow D}) \hspace{1cm}\mbox{and}\hspace{1cm}
({\overrightarrow D}\Psi_{L})\Psi_{R}-\Psi_{L}(\Psi_{R}{\overleftarrow D}) 
\end{eqnarray*}
which in turn suggests we also write the time derivatives as
\begin{eqnarray*}
(\partial_{t}\Psi_{L})\Psi_{R}+\Psi_{L}(\partial_{t}\Psi_{R})
\hspace{1cm}\mbox{and}\hspace{1cm}
(\partial_{t}\Psi_{L})\Psi_{R}-\Psi_{L}(\partial_{t}\Psi_{R})
\end{eqnarray*}
Now the dynamics must include the Dirac derivatives, the external potentials and the mass of the particle, so we will introduce two `types' of Hamiltonian, $\overrightarrow{H}=\overrightarrow{H}(\overrightarrow{D}, V, m)$ and $\overleftarrow{H}=\overleftarrow{H}(\overleftarrow{D}, V, m)$.
Our defining dynamical equations now read
\begin{eqnarray}
i[(\partial_{t}\Psi_{L})\Psi_{R}+\Psi_{L}(\partial_{t}\Psi_{R})]=i\partial_{t}\hat{\rho}=(\overrightarrow{H}\Psi_{L})\Psi_{R}-\Psi_{L}(\Psi_{R}\overleftarrow{H})
				\label{eq:conprob}
\end{eqnarray}
and
\begin{eqnarray*}
i[(\partial_{t}\Psi_{L})\Psi_{R}-\Psi_{L}(\partial_{t}\Psi_{R})]=(\overrightarrow{H}\Psi_{L})\Psi_{R}+\Psi_{L}(\Psi_{R}\overleftarrow{H})
					\label{eq:anticom}
\end{eqnarray*}
The first of these equations (\ref{eq:conprob}) can be written in the more suggestive form
\begin{eqnarray}
i\partial_{t}\hat{\rho}=[H,\hat{\rho}]_{-}
				\label{eq:conprob2}
\end{eqnarray}
This equation has the form of Liouville's equation and can be shown to correspond exactly to the conservation of probability equation. 

Equation (\ref{eq:anticom}) can be written in the form
\begin{eqnarray}
i[(\partial_{t}\Psi_{L})\Psi_{R}-\Psi_{L}(\partial_{t}\Psi_{R})]=[H,\hat{\rho}]_{+}
					\label{eq:anticom2}
\end{eqnarray}
As far as I am aware this equation has not appeared in this form in the literature before it was introduced by Brown and Hiley \cite{mbbh00}.  They arrived at this equation using a different method and showed it was in fact the quantum Hamilton-Jacobi Equation that appears in Bohm's approach to quantum mechanics.  

To bring out exactly what these equations mean in the approach we have adopted here, let us look at their form in the Schr\"{o}dinger Clifford algebra ${\cal{C}}_{0,1}$.  Here $\Psi_{L}=a({\bf r},t) +eb({\bf r},t)$ where $e$ is the generator of the Clifford algebra ${\cal{C}}_{0,1}$ and $a, b \in \Re$.  Similarly $\Psi_{R}=a({\bf r},t) -eb({\bf r},t)$.  This means the $\Psi_{L}\Psi_{R}=a^{2} +b^{2}$ which is just the probability density.  Using $H=p^{2}/2m+V$ it is not difficult to show equation (\ref{eq:conprob2}) becomes
\begin{eqnarray*}
\partial_{t}P+\nabla(P\nabla S/m)=0
\end{eqnarray*}
which is clearly the usual conservation of probability equation.

Equation (\ref{eq:anticom2}) can be considerably simplified if we write $\Psi_{L}=R\exp[eS]$, then after some work we find the equation becomes the quantum Hamilton-Jacobi equation
\begin{eqnarray*}
\partial_{t}S+(\nabla S)^{2}/2m +Q + V =0
\end{eqnarray*}
where $Q=-\nabla^{2}R/2mR$ is the usual expression for the quantum potential.

We have thus arrived at the two dynamical equations that form the basis of the Bohm approach to quantum mechanics.  By generalizing this to the Pauli and Dirac Clifford algebras on can obtain a complete description of the quantum dynamical equations for the Pauli and Dirac particles.  In each case we get a Louville type conservation of probability equation and a quantum Hamilton-Jacobi equation. Thus we find the defining equations used in the Bohm approach appear in each of the three Clifford algebras considered in this paper.  This means we have an approach to relativistic particle quantum mechanics that is completely analogous to the one used by Bohm for the non-relativistic Schr\"{o}dinger equation.  Thus the common misconception that the Bohm approach cannot be applied in the relativistic domain is just not correct.  The full details of our approach to the relativistic domain will be found in Hiley and Callaghan \cite{bhrc08b}

The method we have outlined in this paper improves considerably on the previous attempts to extend the Bohm approach to spin and to the relativistic domain.  For example 
the Pauli equation has already been treated by Bohm, Schiller, and Tiomno, \cite{dbrsjt} and Hestenes, and Gurtler, \cite{dhrg71}, while the Dirac theory has been only been partially treated in terms of the Bohm approach by Bohm and Hiley \cite{dbbh93} and in Hestenes \cite{dh03} our method presented here gives a systematic approach arising directly from connections in the appropriate  Clifford bundles.


\section{Expectation Values.}

\subsection{How to find the coefficients in $\hat{\rho}$.}

In the previous section we have shown how the state of the system is described by the Clifford density element,  $\hat{\rho}= \Psi_{L}\Psi_{R}$, evolves in time.  However we need to extract information about the properties of the quantum system from $\hat{\rho}$.  To do this we need examine the details of $\hat{\rho}$. In general this element takes the form
\begin{eqnarray}
\hat{\rho}=\rho^{s} + \sum \rho^{i}e_{i} + \sum\rho^{ij}e_{ij} + \dots  \nonumber
\end{eqnarray}
Clearly the coefficients, $\rho^{\dots i\dots}$, characterise different aspects of the state of the system.  How then do we abstract these coefficients from the general expression? 

Notice first that any Clifford element, $B$,  can also be written in the form
\begin{eqnarray}
B = b^{s} + \sum b^{i}e_{i} + \sum b^{ij}e_{ij} + \dots \nonumber
\end{eqnarray}
Secondly,  all Clifford algebras contain a notion of trace, where the trace of the identity element is the dimension of the ${\cal{I}}_{L_{m}}$ and the trace of each $e_{A}$ is zero except when $A = 0$. Then
\begin{eqnarray}
tr(B) = b^{s}n  \nonumber
\end{eqnarray}
Thus the scalar coefficient corresponds to taking the trace.  To find the other coefficients we note that if we  multiply $B$ by the appropriate basis element $e_{\dots}$, we will always be left with a scalar term.  This term will then be the trace.  But of course this will just give the coefficient that stands in front of the particular $e_{\dots}$ chosen.  For example if we want to find $b^{k}$, we form $tr(Be_{k})$ which then picks out $b^{k}$ because
\begin{eqnarray}
tr(Be_{k})=nb^{k} \nonumber
\end{eqnarray}
In this way we can systematically find all the coefficients of any element in the algebra.

To illustrate this in the particular case of the Dirac Clifford, we write 
\begin{eqnarray}
B = b^{s} + \sum b^{i}e_{i} + \sum b^{ij}e_{ij} + \sum b^{ijk}e_{ijk} + b^{5}e_{5}  \nonumber
\end{eqnarray}
 Then we have 5 trace-types
\begin{eqnarray}
b^{s}=tr(B1)/4 \hspace{1cm}b^{i}= tr(Be_{i})/4\hspace{1cm}b^{ij}= - tr(Be_{ij})/4\nonumber\\
b^{ijk}= - tr(Be_{ijk})/4\hspace{1cm}b^{5}= tr(Be_{5})/4\hspace{2cm} \nonumber
\end{eqnarray}
Thus we can find all the coefficients of any element of the Dirac Clifford.


\subsection{How to Find Expectation Values.}

In standard quantum mechanics, expectation values of operators are found via:
\begin{eqnarray}
\langle B \rangle =\langle \psi |B| \psi \rangle =tr(B\rho) \nonumber
\end{eqnarray}
Here $\rho$ is the standard density matrix.
These mean values can be taken over directly into the algebraic approach by replacing the standard density matrix by the $\hat{\rho}$ as can easily be seen from the following correspondences:-
\begin{eqnarray}
|\psi\rangle \rightarrow\Psi_{L}=\psi_{L}\epsilon\hspace{2cm}\langle\psi|\rightarrow \Psi_{R}=\epsilon\psi_{R} \nonumber
\end{eqnarray}
Then we can write
\begin{eqnarray}
\langle B\rangle = tr(\epsilon\psi_{R} B \psi_{L}\epsilon) \nonumber
\end{eqnarray}
Since the trace is invariant under cyclic permutations, we can arrive at the form
\begin{eqnarray}
\langle B\rangle = tr(\psi_{R} B \psi_{L}\epsilon)= tr(B\psi_{L}\epsilon\psi_{R}) \label{eq:expt}
\end{eqnarray}
But $ \psi_{L}\epsilon\psi_{R}=\hat{\rho}$, so we have
\begin{eqnarray}
\langle B\rangle = tr(B\hat{\rho}) \nonumber
\end{eqnarray}
Thus the mean value of any dynamical element, $B$, in the Clifford algebra becomes
\begin{eqnarray}
tr(B\hat{\rho}) = tr(b^{s}\hat{\rho} + \sum b^{i}e_{i}\hat{\rho} +\sum b^{ij}e_{ij}\hat{\rho} +\dots)\hspace{1.3cm}  \nonumber\\
= b^{s}tr(\hat{\rho}) + \sum b^{i}tr(e_{i}\hat{\rho}) + \sum b^{ij}tr(e_{ij}\hat{\rho})/2 \dots\nonumber
\end{eqnarray}
This shows that the state of our system is specified by a set of bilinear invariants
\begin{eqnarray}
tr(1\hat{\rho})\rightarrow\mbox{scalar}\hspace{0.7cm}tr(e_{j}\hat{\rho})\rightarrow\mbox{vector}\hspace{0.7cm}tr(e_{ij}\hat{\rho})\rightarrow\mbox{bivector}\hspace{0.7cm}tr(\dots)\rightarrow\dots\hspace{0.7cm} \nonumber 
\end{eqnarray}
These bilinear invariants play an important role in  the algebraic approach.


\subsection{Some Specific Expectation Values.}

We now give a specific example of how we calculate expectation values  in the Pauli Clifford algebra. Suppose we require the expectation value of a component of the spin of a Pauli particle.  We identify the spin components with the elements $\{e_{i}\}$ of the algebra. Then, specifically for $i = j $, we find that the expectation value of the $j$th component of the spin is given by
\begin{eqnarray}
\langle S_{j}\rangle =tr(e_{j}\hat{\rho}).  \label{eq:mean}
\end{eqnarray}
In order to evaluate $\hat{\rho}=\psi_{L}\epsilon\psi_{R}$, we need to chose a specific expression for $\epsilon$.  Since it is conventional to choose the $z$-direction as the direction of any applied homogeneous magnetic field, we choose for our idempotent $(1 + e_{3})/2$. Thus
\begin{eqnarray}
2\langle S_{j}\rangle = tr(e_{j}\psi_{L}(1 + e_{3})\psi_{R})= tr(e_{j}\psi_{L}e_{3}\psi_{R}).  \nonumber
\end{eqnarray}
Let us define $\rho S =\psi_{L}e_{3}\psi_{R}$ where $\Psi_{L}$ is given by
\begin{eqnarray}
\Psi_{L}=\psi_{L}\epsilon=(a + e_{123}d)(1 + e_{3})/2 + (c +  e_{123}b)(e_{1} + e_{13})/2 \label{eq:PLI}
\end{eqnarray}
Here $a, b, c$ and $d$ are related to the components of the usual matrix representation (See equation \ref{eq:matspin1})	
\begin{eqnarray}
\psi=\begin{pmatrix}
  \psi_{1} \\
  \psi_{2}
 \end{pmatrix}
				\label{eq:matspin}			
\end{eqnarray}
via
  \begin{eqnarray}
2a= \psi_{1} + \psi^{*}_{1}\hspace{2.7cm}2c=\psi_{2} + \psi^{*}_{2}\hspace{0.4cm}\nonumber\\
2e_{123}d=  \psi_{1} - \psi^{*}_{1}\hspace{2cm}2e_{123}b=\psi_{2} - \psi^{*}_{2}. \label{eq:PLIWave}
\end{eqnarray}  
Using equations (\ref{eq:PLI}) and (\ref{eq:PLIWave}), it can be shown that $S$ is the usual expression for the spin vector
\begin{eqnarray}
\rho S=(\psi_{1}\psi^{*}_{2} + \psi_{2}\psi^{*}_{1})e_{1} +i(\psi_{1}\psi^{*}_{2} - \psi_{2}\psi^{*}_{1})e_{2} + (|\psi_{1}|^{2}|-|\psi_{2}|^{2})e_{3}  \label{eq:spin}
\end{eqnarray}
Thus we find
\begin{eqnarray}
\langle S_{j}\rangle = tr(e_{j}\rho S) = (e_{j}\cdot \rho S)/2  \nonumber
\end{eqnarray}
Here we use the conventional notation that $(A\cdot B)/2$ is the scalar part of the Clifford product.  This simply picks out the appropriate coefficient of $e_{j}$ in equation (\ref{eq:spin}).  Thus for example, the mean value of the $x$-component of the spin is 
\begin{eqnarray}
\langle S_{1}\rangle = (\psi_{1}\psi^{*}_{2} + \psi_{2}\psi^{*}_{1})/2. \nonumber
\end{eqnarray}
If the particle is in a state of spin `up' along the 3-axis,  then we find
\begin{eqnarray}
\langle S_{1}\rangle =\langle S_{2}\rangle =0\hspace{1cm}\langle S_{3}\rangle=1/2 \nonumber
\end{eqnarray}
If, on the other hand spin is down along the 3-axis we find
\begin{eqnarray}
\langle S_{1}\rangle =\langle S_{2}\rangle =0\hspace{1cm}\langle S_{3}\rangle=- 1/2 \nonumber
\end{eqnarray}
Thus we see how the eigenvalues of a given physical element like spin are identified in the Clifford algebra approach without the need to bring in operators working in a Hilbert space.
This whole process can be generalised to any element of any Clifford algebra.  We will not discuss the details of this generalisation here.


\subsection{Mean Values of Differential Elements.}

We will assume the elements of our ${\cal{I}}_{L_{m}}$s  to be functions of the co-ordinates of the underlying manifold, therefore as we have seen above differentiation is possible.  In order to continue making contact with standard quantum mechanics, we need to introduce elements like $E=B\partial/\partial t$ and $P=B\partial/\partial x_{i}$, where $B$ is some general element of the Clifford algebra that has physical significance.  Since we have to consider both differentiation from the left and the right, our expectation values take the form
\begin{eqnarray}
2\langle P\rangle_{\pm} = tr\left[\epsilon\psi_{R}(P \psi_{L})\epsilon\right]
\pm tr{\left[\epsilon\psi_{L}(\tilde{P}\psi_{R}) \epsilon\right]} \label{eq:expt2}
\end{eqnarray}
 Now let us write $P = B\partial$ where $B$ is again any element of the Clifford algebra. Then
\begin{eqnarray}
2\langle P\rangle_{\pm} =tr\left[B(\partial\psi_{L})\epsilon\psi_{R}\right]
\pm tr\left[\tilde{B} \psi_{L}\epsilon(\partial\psi_{R})\right].  \end{eqnarray}
To check the consistency of our approach to expectation values let us suppress the differential, $\partial$ in equation \eqref{eq:expt2} 
and simply write $P = B$ 
\begin{eqnarray}
2\langle P\rangle_{\pm} =tr\left[B\psi_{L}\epsilon\psi_{R}\right]
\pm tr\left[\tilde{B} \psi_{L}\epsilon\psi_{R}\right]. \label{eq:MV}
\end{eqnarray}
This expression reduces clearly reduces to equation (\ref{eq:mean}) if $B = \tilde{B}$ which is what is desired for consistency. 


\subsection{Simple Example.}

	We will now give an example to how this approach works in two simple cases.  Let us find  bilinear invariant for the energy in the Schr\"{o}dinger case.  We start from the Clifford  energy element $e\partial_{t}$ and we evaluate the energy $\langle \hat{E}\rangle_{+} \equiv \rho E$.  Thus
	\begin{eqnarray}
2\rho E = e[(\partial_{t}\psi_{L})\psi_{R} - \psi_{L}(\partial_{t}\psi_{R})]  \label{eq:ES1}
\end{eqnarray}
Here we have written $\epsilon = 1$ as this is the only idempotent in the Schr\"{o}dinger Clifford algebra. In this algebra 
\begin{eqnarray}
\Psi_{L}=a+eb\hspace{1cm}\mbox{while}\hspace{1cm}\Psi_{R}=a-eb			\label{eq:SLI}
\end{eqnarray}
where $a$ and $b$ are related to the conventional wave function via
\begin{eqnarray}
2a=\psi+\psi^{\dag}\hspace{1cm} \mbox{and}\hspace{1cm} 2b=\psi-\psi^{\dag}			\label{eq:SLIW}
\end{eqnarray}

 Substituting equation  \eqref{eq:SLI}  into \eqref{eq:ES1} gives
\begin{eqnarray}
\rho E = (\partial_{t} a) b - a(\partial _{t} b).  \nonumber
\end{eqnarray}
and then using the relations between the real parameters and the standard wave functions given in equation (\ref{eq:SLIW}), we find
\begin{eqnarray}
2\rho E = e[(\partial_{t}\psi)\psi^{*}- (\partial _{t}\psi^{*})\psi ] \nonumber
\end{eqnarray}
Finally if we write $\psi = R \exp[eS]$  we obtain
\begin{eqnarray}
E=- \partial_{t} S.  \label{eq:ES2}
\end{eqnarray}
Those familiar with the Bohm model will immediately recognize that this as the Bohm energy.  Indeed if we work out the corresponding expression for the Clifford momentum element $ \hat{P} = e\nabla $ we find it gives the Bohm momentum.  To show this, substitute this expression into equation (\ref{eq:MV}) and writing $\langle \hat{P}\rangle_{+} = \rho{\bf P}$, we find ${\bf P} = \nabla S$. This will immediately be recognized as the Bohm momentum.  

Thus it appears as if the Bohm model literally drops out of the Clifford algebra approach.  Notice also that we are using the Clifford algebra over the reals, with the Clifford element $e$ playing the role of $i$.  We have discussed this consequence and all that follows from it in Hiley and Callaghan \cite{bhrc08a} .  This method has been applied to Pauli Clifford in \cite{bhrc08a} and  the Dirac Clifford in  \cite{bhrc08b}.

\section{The Symplectic Group.}

So far we have concentrated on orthogonal properties  and we have shown how the  formulation using the notion of {\em process} leads to an orthogonal groupoid and then onto an orthogonal Clifford algebra.  The rotations were handled using the  Clifford group which operates on general elements of the Clifford algebra  through  inner automorphisms.  This enabled us to construct the light cone structure of space-time.  

We introduced the dynamical structure by exploiting the structure of the connections on a Clifford spin bundle.  This actually breaks with the main aim of this work, namely to capture the underlying process completely in algebraic terms.  Rather than capture the dynamical structure using connections we now  introduce the symplectic groupoid.  Here we are exploiting a suggestion of Connes \cite{ac} who has already pointed out that Heisenberg's earlier attempts to create quantum mechanics exploits this symplectic groupoid structure.

We will follow in the same spirit by considering how we go from a space-time structure to what is essentially a phase space structure. To this end we now introduce a collection of qualitatively new movements that will double the algebraic structure we are exploring. Therefore we distinguish two types of movement.  A space-time movement which we will now denote by $X[P_{n}P_{m}]$ and a new movement $P[P_{n}P_{m}]$ which will correspond to a movement in what we would conventionally be called the momentum space. However recall we start with no {\em a priori} given underlying manifold.  It is intention to generate any underlying manifolds from the algebra of process itself.  Thus in this notation  $X$ and $P$ simply label the qualities of movement.  We then introduce the product $X[P_{n}P_{k}]\bullet P[P_{k}P_{m}]=XP[P_{n}P_{m}]$. We will assume that this product is not commutative and follow Heisenberg by assuming it satisfies the commutator
\begin{eqnarray*}
X[P_{n}P_{k}]\bullet P[P_{k}P_{n}]-P[P_{n}P_{k}]\bullet X[P_{k}P_{n}]=i=[X,P]
\hspace{1cm}(\hbar=1)
\end{eqnarray*}
In this way we arrive at a structure which has similarities with the structure used by Dirac \cite{pd26} in his discussions of the quantum algebra. One of the technical problems of using the symplectic groupoid in the above form is that it is  an infinite dimensional algebra and therefore it is difficult to see what is going on.  We have found it much easier to illuminate the structure if we consider instead, the finite Weyl algebra, $C^{2}_{n}$ \cite{am67}, where $n$ is an integer $0< n\leq \infty$.  Then we have
\begin{eqnarray*}
X[P_{n}P_{k}]\bullet P[P_{k}P_{n}]=\omega P[P_{n}P_{k}]\bullet X[P_{k}P_{n}]
\end{eqnarray*}
Here $\omega^{n} = 1$ so that $\omega $ is the $n^{\mbox{th}}$ root of unity.  We can simplify the notation by writing $X[P_{n}P_{k}]=U$ and $P[P_{k}P_{n}]=V$ so that our algebra becomes
\begin{eqnarray*}
UV =\omega VU;\qquad U^{n} =1 \qquad V^{n}=1.
\end{eqnarray*}
We will show that as $n \rightarrow \infty$, the discrete Weyl algebra becomes continuous and is essentially the Heisenberg algebra but with a small but significant difference.  The discrete Weyl algebra allows us to explore what we will call a toy phase space.

The discrete Weyl algebra $C^{2}_{n}$ contains a set of idempotents, $\epsilon_{j}$.  We will choose a particular set given by 
\begin{eqnarray*}
\epsilon_{j} = \frac{1}{n}\sum_{k}\omega^{-jk}R(0,k)\hspace{1cm}\mbox{where}\hspace{1cm}R(j,k)=\omega^{-jk/2}U^{j}V^{k},\quad j,k=0,1,\dots ,n-1
\end{eqnarray*}
These idempotents satisfy the relation $\sum_{k}\epsilon_{k}\epsilon_{j}=\epsilon_{j}$.  In keeping with the spirit we have been developing in this paper, we will regard the idempotents as `points' which we can map onto some underlying vector space.

If we now write $U=T=\exp\left[2\pi \delta xP/n\right]$ then we can generate a set of points using the inner automorphisms 
\begin{eqnarray*}
\epsilon_{j+1}=T\epsilon_{j}T^{-1}
\end{eqnarray*}
In this way we generate a set of points which we display in figure 8 below.

\begin{figure}[h]
\begin{center}
\includegraphics[width=3in]{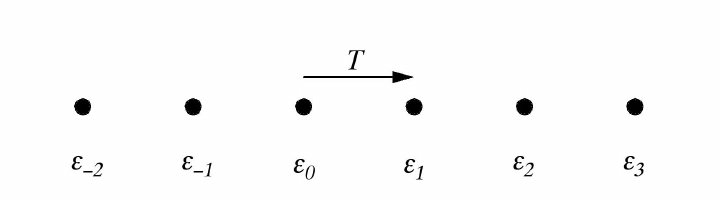}
\caption{Set of `space' points generated by $T$.  }
\end{center}
\end{figure}
We can introduce an element of the algebra, $X=\delta x\sum_{k}\epsilon_{k}$ so that $X\epsilon_{j}=\delta xj\epsilon _{j}$.  In other words the element $X$ plays the role of a position operator enabling us to label the point with $j$.  It should be noted that $\delta j \epsilon_{j}=\Psi_{L}$ is an element of a minimal left ideal.

We can also form a different set of idempotents 
\begin{eqnarray*}
\epsilon'_{j}=\frac{1}{n}\sum_{k}\omega^{-jk}R(k,0)
\end{eqnarray*}
This enables us to generate a set of other points using $V=T'=\exp\left[2\pi \delta pX/n\right]$
\begin{figure}[h]
\begin{center}
\includegraphics[width=3in]{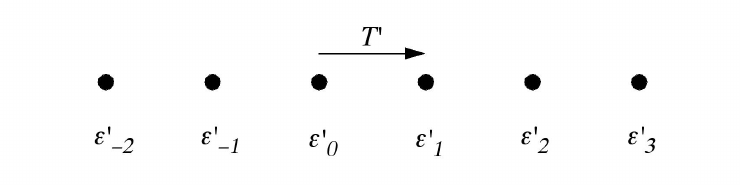}
\caption{Set of `momentum' points  }
\end{center}
\end{figure}
We can again introduce an element of the algebra $P=\delta p\sum_{j}j\epsilon'_{j}$ so that $P\epsilon'_{j}=\delta pj\epsilon'_{j}$.  Thus $P$ enables us to label the momentum points by, say, $j_{p}$.  Again it should be noted that $\delta pj\epsilon'_{j}=\Phi_{L}$ is an element of another minimal left ideal.

In other words we have generated a pair of complementary spaces, the position space and the momentum space.  One important lesson we learn from this model is that it is not possible to generate both spaces simultaneously.  This is because the two sets of idempotents are related by
\begin{eqnarray*}
\epsilon'_{j} = Z^{-1}\epsilon_{j}Z\hspace{1cm}\mbox{with}\hspace{1cm} Z=\frac{1}{\sqrt{n^{3}}}\sum_{ijk}\omega^{j(i-k)}R(j-i,k)
\end{eqnarray*}

\begin{figure}[h]
\begin{center}
\includegraphics[width=2in]{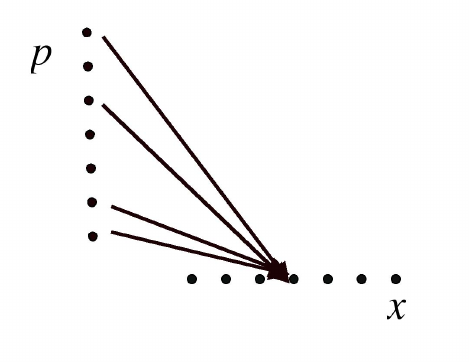}
\caption{Many $p$-points are enfolded into each $x$-point.   }
\end{center}
\end{figure}

This transformation is, of course, a finite Fourier transformation.  If this approach is correct, then it shows that the origins of the uncertainty principle is not that we disturb the phenomena with our `blunt' experimental instruments, but that the nature of physical processes themselves are such that it is structurally impossible to display position and momentum simultaneously using algebraic idempotents.  

In fact each finite Fourier transformation `explodes' every space point into every momentum point and {\em vice versa}.  This means that the $p$-points are not `hidden', they are not manifest. We can regard them as {\em enfolded} (see figure 10.). Essentially what we have is that the $p$-points are `hologramed' into the $x$-points and {\em vice versa}.  In this way we have, as it were, an ontological principle of complementarity which is to be contrasted with Bohr's epistemic principle of complementarity.   A more detailed discussion of the mathematics lying behind this approach will be found in Hiley and Monk \cite{bhnm93})

\subsection{The Continuum Limit.}

In this section we want to show how to proceed to the continuum limit essentially following the procedure outlined by Weyl \cite{hw}.  To this end let us simplify the notation and write
\begin{eqnarray*}
\xi=\frac{2\pi \delta x}{n}\hspace{1cm}\mbox{and}\hspace{1cm}
\eta=\frac{2\pi \delta p}{n}
\end{eqnarray*}
Then we have
\begin{eqnarray*}
U=\exp[i\xi P]\qquad V=\exp[i\eta X]\qquad R(j,k)=\omega^{-jk/2}U^{j}V^{k}
\end{eqnarray*}
with $\omega=\exp[i\xi\eta]$.  In this case the idempotent becomes
\begin{eqnarray*}
\epsilon_{0}=\frac{1}{n}\sum_{k}R(0,k)=\frac{1}{n}\sum_{k}V^{k}=\frac{1}{n}\sum_{k}\exp[ik\eta X]
\end{eqnarray*}
Now $k\eta$ runs through all integer values, but as $\eta \propto 1/n$ and $n\rightarrow \infty$ $k\eta$ may assume all the real numbers from $-\infty$ to $+\infty$.  In this case
\begin{eqnarray*}
\epsilon_{0}=\frac{1}{n}\sum_{k}\exp[ik\eta X]\rightarrow \frac{1}{2\pi}\int d\beta \exp[i\beta q] =\delta(q)
\end{eqnarray*}
If we now form
\begin{eqnarray*}
\epsilon_{\xi}=\exp[-i\xi P]\epsilon_{0}\exp[i\xi P]=\frac{1}{n}\sum_{\beta}\exp[i\xi P]\exp[i\beta X]\exp[i\xi P]=\frac{1}{n}\sum_{\beta}\exp[i\beta(X-\xi)]\\
\rightarrow\frac{1}{2\pi}\int d\beta \exp[i\beta(X-\xi)=\delta(q-\xi)\hspace{4cm}
\end{eqnarray*}
In this way we see the idempotents become delta functions so we have generated the continuum of points.  We can now choose another set of points from $\epsilon'_{0}$ using $U$ to translate between points so that we can generate another set of points, the $p$-continuum.  In this way we see that the lesson we learned in the discrete case generalises to the continuum.

\subsection{The Schr\"{o}dinger Representation}

Let us take this a little further and  write $\delta xj\epsilon_{j}=x_{j}$ so that $x_{j} \in {\cal{I}}_{L}$. Then
\begin{eqnarray*}
U^{s}:x_{j}\rightarrow x_{j-s}\hspace{1cm} V^{t}:x_{j}\rightarrow \omega^{jt}x_{j}
\end{eqnarray*}
Now let us consider the limit $n \rightarrow \infty$, then 
\begin{eqnarray*}
s\xi\rightarrow\sigma\qquad t\eta \rightarrow\tau \qquad \omega^{kt}=\exp[i\xi kt\eta]=\exp[iq\tau];\quad(k\xi=q)
\end{eqnarray*}
Here $k$ is an integer, but $\xi \propto 1/n$ which means that for $n$ large, \hspace{0.1cm}$k$ runs from $- \infty$ to $+\infty$.  This is because here $k$ is considered as mod $n$, while \ $k\xi$ is mod $n\xi$, but $n\xi$ is a multiple of $2\pi/\eta$ so that $\pi/\eta \rightarrow \infty$ as $\eta \rightarrow 0$.  In this case $x_{j}$ becomes continuous variable $X$ through the relation $x_{j}=\sqrt{\xi}X$.  Then
\begin{eqnarray*}
U^{s}:X\rightarrow\exp[i\sigma P]X\rightarrow(X-\sigma)
\qquad V^{t}:X\rightarrow \exp[i\tau X]X.
\end{eqnarray*}
These relationships are in the symplectic Clifford algebra.
It should immediately be recognised that this is one short step from the Schr\"{o}dinger representation.  In other words we have arrived at the usual continuum limit.

There is more work to be done to smoothly join the symplectic structure to the orthogonal Clifford algebra.  Some of the mathematical details have been explored in  Crumeyrolle \cite{ac90} but space limits further discussion.

\section{General Conclusions.}

In this article we have shown that by starting from a basic primitive notion of process, activity or movement, we have generated both orthogonal and symplectic Clifford algebras.  We have shown that the orthogonal structure enables us to discuss the structure of space-time in terms of a light ray geometry.  We found that this structure is entirely  classical even though we found ourselves using the Pauli and Dirac Clifford algebras, normally assumed to be quantum in essence.

To introduce quantum aspects into our description, we used two approaches.  The first exploited connections on a Clifford bundle.  This led to two specific forms of connection, one operating from the left and the other from the right acting on what is the algebraic equivalent of the density matrix. However our methods are independent of any specific matrix representation.  The resulting dynamical equations were equivalent to the two dynamical equations exploited by the Bohm approach.  This shows that the Bohm approach is not some peripheral structure in quantum mechanics but an essential part of standard algebraic quantum mechanics when discussed in terms of Clifford algebras.  

We found the Schr\"{o}dinger, Pauli and Dirac theories formed a  hierarchy of Clifford algebras, each fitting naturally inside the other corresponding to the physical hierarchy, non-relativistic particle without spin, non-relativistic particle with spin and relativistic particle with spin. This provides a natural description for quantum phenomena, not in terms of representations in Hilbert space, although such a representation is available if desired, but in terms of Clifford algebras in which specific representations are not required.

The appearance of the defining equations of the Bohm approach provides a clear extension of the approach to the relativistic domain showing the criticism that the approach fails in the relativistic domain unfounded.  The Dirac theory produces a `quantum potential' which can be shown to be the exact relativistic generalisation of the quantum potential found originally by de Broglie \cite{ldb} and Bohm \cite{db52}.  Space does not allow us to discuss these details in this paper but will be reported elsewhere in Hiley and Callaghan \cite{bhrc08a}, \cite{bhrc08b}.

In the final section we generalised the process structure by introducing a symplectic groupoid.  We did this by using a discrete structure which in the limit approached the Heisenberg algebra.  This algebra led to a general non-commutative structure in which it was not possible to create points in position and momentum simultaneously.  This provides a clear ontological basis for what Bohr called the principle of complementarity. This algebraic structure provides the mathematical structure that lies behind Bohm's notion of the implicate order.  The projections, the shadow manifolds are examples of the explicate order.  The simple original Bohm model of quantum mechanics simply produces a phase space example of one of these shadow manifolds.  Thus we have provided a more general mathematical framework within which it is possible to further explore the quantum world.

\section*{Acknowledgements}

I should like to thank in particular Bob Callaghan for his patience during the many discussions we had on various aspects of this subject.  I would also like to thank Ernst Binz, Ray Brummelhuis,  Bob Coecke, Maurice de Gosson and Clive Kilmister  for their continual encouragement.  Without Ray Brummelhuis' generous support this work would not have reached the light of day.




\begin{thebibliography}{99}

\bibitem{nb61} Bohr, N.,   {\em Atomic Physics and Human Knowledge}, Science Editions, New York, (1961).

\bibitem{aw} Whitehead, A. N.,   {\em Process and Reality},  Harper \& Row, New York, (1957).


\bibitem{db86} Bohm, D., Time, the Implicate Order and Pre-Space, in {\em Physics and the Ultimate Significance of Time}, ed. D.R. Griffin, 172-6 \& 177-208,  SUNY Press, N.Y., (1986).

\bibitem{dbpdbh}Bohm, D. J., Davies, P. G., and Hiley, B. J.,1982, Algebraic Quantum Mechanics and Pre-geometry, {\em Quantum Theory: Reconsiderations of FoundationsÑ3}, V\"{a}xj\"{o} , Sweden 2005, ed Adenier, G., Krennikov, A. Yu. and Nieuwenhuizen, Theo. M., pp. 314-24, AIP, (2006).


\bibitem{bdbh84}Bohm, D., and Hiley, B. J.,  Generalization of the Twistor to Clifford Algebras as a basis for Geometry,  {\em Rev. Briasileira de Fisica, Volume Especial, Os 70 anos de Mario Sch\"{o}nberg},  (1984) 1-26

\bibitem{gg} Grassmann, G. {\em A New Branch of Mathematics: the Ausdehnungslehre of 1844 and Other Works}, translated by Kannenberg, L. C., Open Court, Chicago, (1995).


\bibitem{wh67} Hamilton, W.R., {\em Mathematical Papers}, Vol. 3 Algebra, Cambridge (1967)

\bibitem{dh03} Hestenes, D., Spacetime Physics with Geometric Algebra, {\em Am. J. Phys}, {\bf 71}, (2003), 691-704,  

\bibitem{cdal} Doran, C. and Lasenby, A.,  {\em Geometric algebra for physicists}, Cambridge University Press, Cambridge , 2003.

\bibitem{gg3} Grassmann, G., {\em ibid}. section 3 and 4.

\bibitem{jdmh} Demaret, J., Heller, M. and Lambert, D., Local and Global Properties of the world, {\em Foundations of Science}, {\bf 2}, 137-176, (1997)

\bibitem{lk98} Kauffman, L. H., Space and Time in Computation and Discrete Physics, {\em Int. J. General Systems,} {\bf 27}, (1998), 249-273.

\bibitem{db76} Bohm, D., {\em Fragmentation and Wholeness}, The van Leer Jerusalem Foundation, Jerusalem, (1976)

\bibitem{dbbhas70} Bohm,D., Hiley, B. J. and  Stuart,A. E. G., On a New Mode of Description in Physics,  {\em Int. J. Theor. Phys}., {\bf 3}, (1970), 171-183. 

\bibitem{hb26} H. Brandt, H., Uber eine Verallgemeinerung des Gruppenbegriffes, {\em Math. Ann}. {\bf 96} (1926) 360-366.

\bibitem{db65} Bohm, D.,  Space, Time, and the Quantum Theory Understood in Terms of Discrete Structural Process, {\em Proc. Int. Conf. on Elementary Particles}, Kyoto, 252-287, (1965)

\bibitem{ae58}  Eddington, A. S.,   {\em The Philosophy of Physical Science},  Ann Arbor Paperback, University of Michigan Press, Michigan, (1958).

\bibitem{pl97} Lounesto, P., {\em Clifford Algebras and Spinors}, Cambridge University Press, Cambridge, (1997).


\bibitem{rp67} Penrose, R.,  Twistor Algebra,  {\em J. Maths Phys}., {\bf 8}, (1967), 345-366.

\bibitem{ffbh80} Frescura, F. A. M. and Hiley, B. J.,   The Implicate Order, Algebras, and the Spinor,  {\em Found. Phys.}, {\bf 10}, (1980), 7-31.

\bibitem{bhnm93} Hiley, B. J. and  Monk, N., Quantum Phase Space and the Discrete Weyl Algebra, {\em Mod. Phys. Lett.}, {\bf A8}, (1993), 3225-33.

\bibitem{lk87} Kauffman, l. H.,  Special Relativity and a Calculus of Distinctions, in {\em Proc. 9th Int. Meeting of ANPA}, pp. 291-312, Cambridge, ANPA West, Palo Alto, Cal., (1987).

\bibitem{ag00} Griffor, A., From Strings to Clifford Algebras, in {\em Implications: Scientific Aspects of AHPA 22.}  ed., K. G. Bowden, pp. 30-55, 2001.

\bibitem{rb87}   Brown, R.,  From Groups to Groupoids: A Brief Survey, {\em London Math. Soc}. {\bf 19} (1987) 113-134.

\bibitem{sabc04}  Abramsky and Coecke q-ph/0402130

\bibitem{jrz} Raptis, I., Zapatrin, R. R., Algebraic description of spacetime foam. {\em Classical and Quantum Gravity}, {\bf 20} (2001) 4187Ð4205, gr-qc/9904079 

\bibitem{ms57} Sch\"{o}nberg, M., (1957) Quantum Kinematics and Geometry, {\em Nuovo Cimento, Supp}. {\bf VI}, (1957), 356-380.

\bibitem{mf95} Fernandes, M., {\em Geometric Algebras And The Foundations Of Quantum Theory}. PhD Thesis, London University, (1995).

\bibitem{lk82}  Kauffman, K. L., Sign and Space, in {\em In Religious Experience and Scientific Paradigms}, Proc. of the IAWSR Conf., Inst, Adv. Stud. World Religions, Stony Brook, New York, 118-164, (1982).

\bibitem{br79}  Bratteli, O. and Robinson, D. W.,Operator Algebras
and Quantum StatisticaIMechanics I: C*- and W*-Algebras Symmetry Groups Decomposition of States, Springer, Berlin,  (1987).

\bibitem{df96} Finkelstein, D. R., {\em Quantum Relativity: a Synthesis of the ideas of Einstein and Heisenberg}, Springer, Berlin, (1996).

\bibitem{rz01}  Zapatrin, R. R., Incidence algebras of simplicial complexes, {\em Pure Mathematics and its Applications}, {\bf 11}, (2001) 105Ð118; eprint math.CO/0001065.

\bibitem{bh91} Hiley, B. J., Vacuum or Holomovement, in {\em The Philosophy of Vacuum}, ed S. Saunders and H. R. Brown, pp. 217-249, Clarendon Press, Oxford, (1991).

\bibitem{rf} Feynman, R. P., Space-time Approach to Non-Relativistic Quantum Mechanics, {\em Rev. Mod. Phys}. {\bf 20}, (1948), 367-387.

\bibitem{sk00} Kauffman, S. A. , {\em Investigations}, Oxford University Press, Oxford, (2000).

\bibitem{bh01}Hiley, B. J., Towards a Dynamics of Moments: The Role of Algebraic Deformation and Inequivalent Vacuum States, {\em Correlations} ed. K. G. Bowden, Proc. ANPA {\bf 23}, (2001), 104-134.

\bibitem{pd74}  Dirac, P. A. M., {\em Spinors in Hilbert Space.}, Plenum Press, New York, (1974).

\bibitem{lk01} Kauffman, L. H., {\em Knots and Physics}, World Scientific, Singapore, (2001) 


\bibitem{bh08a} Hiley, B. J., Some Notes on Quantum Mechanics using Clifford Algebras, (Available in pre-print form.)

\bibitem{bhrc08a} Hiley, B. J. and Callaghan, R., The Clifford Algebra approach to Quantum Mechanics A: The Schr\"{o}dinger and Pauli Particles, (Available in pre-print form.)

\bibitem{bhrc08b} Hiley, B. J. and Callaghan, R., The Clifford Algebra approach to Quantum Mechanics B: The Dirac Particle, (Available in pre-print form.)

\bibitem{rpwr84} Penrose, R. and Rindler, W., {\em Spinors and Space-time}, Volume 1, Cambridge University Press, Cambridge, 1984.

\bibitem{page} Page, L., 
A New Relativity. Paper I. Fundamental Principles and Transformations Between Accelerated Systems, {\em Phys. Rev}. {\bf 49}, (1936), 254 - 268. 

\bibitem{hb} Bondi, H., {\em Relativity and Common Sense}, Dover, New York, (1977) 

\bibitem{mn90} Nakahara, M., {\em Geometry, Topology and Physics}, Adam Hilger, Bristol, (1990).

\bibitem{ffbh81} Frescura, F. A. M. and Hiley, B. J., Geometric Interpretation of the Pauli spinor, {\em Am. J. Phys}., {\bf 49}, (1981), 152-7.

\bibitem{lr85}  Ryder, L. H., {\em Quantum Filed Theory}, Cambridge University Press, Cambridge, 1985.


\bibitem{bh01a}  Hiley, B. J., Non-Commutative Geometry, the Bohm Interpretation and the Mind-Matter Relationship, in {\em Computing Anticipatory Systems: CASYS 2000-Fourth International Conference}, edited by D. M. Dubois, pp. 77-88, AIP, (2001).

\bibitem{db80} Bohm, D., {\em Wholeness and the Implicate Order}, Routledge, London, 1980.

\bibitem{hbl89} Lawson, H. B. and Michelsohn, M-L., {\em Spin Geometry}, Princeton University Press, Princeton, 1989.

\bibitem{mbbh00} Brown, M. R. and Hiley, B. J., Schr\"{o}dinger revisited: an algebraic approach.  quant-ph/0005026.

\bibitem{dbrsjt} Bohm, D., Schiller, R. and Tiomno, J., A Causal Interpretation of the Pauli Equation (A) and (B), {\em Nuovo Cim}. Supp. {\bf 1}, (1955) 48-66 and 67-91.

\bibitem{dhrg71} Hestenes, D., and Gurtler, R.,  Local Observables in Quantum Theory, {\em Am. J. Phys}. {\bf 39}, (1971) 1028-38.

\bibitem{dbbh93} Bohm, D. and Hiley, B. J.,  {\em The Undivided Universe: an Ontological Interpretation of Quantum Theory}, Routledge, London  1993.



\bibitem{ac} Connes, A., {\em Non-commutative Geometry}, Academic Press, San Diego, 1990.

\bibitem{pd26} Dirac, P. A. M.,  The Fundamental Equations of Quantum Mechanics, {\em Proc. Roy. Soc.}, A{\bf 109} (1926) 642-653.

\bibitem{am67} Morris, A. O.,  On a Generalised Clifford Algebra, {\em Quart. J. Math}. {\bf 18}, (1967) 7-12.

\bibitem{bh08} Hiley, B. J., {\em Clifford Algebras}, in preparation.

\bibitem{ck94} Kilmister, C. W., {\em Eddington's search for a fundamental theory}, Cambridge University Press, Cambridge, 1994.

\bibitem{hw}  Weyl, H., {\em The Theory of Groups and Quantum Mechanics} Dover Publication, London, 1960.

\bibitem{ac90}  Crumeyrolle A., {\em Orthogonal and Symplectic Clifford Algebras:  Spinor Structures}, Kluwer, Dordrecht, 1990.

\bibitem{ldb} de Broglie, L., {\em Non-linear Wave Mechanics: A Causal Interpretation.} Elsevier, Amsterdam, 1960.

\bibitem{db52} Bohm, D.,  A Suggested Interpretation of the Quantum Theory in Terms of Hidden Variables, I, {\em Phys. Rev}., {\bf 85}, (1952), 166-179;  and II, {\bf 85}, (1952), 180-193.

\end{thebibliography}
\end{document}